\documentstyle[12pt]{article} 
\newcommand{\calO}{{\cal O}}

\newcommand\exa{\nopagebreak \begin{flushleft}\smallskip \nopagebreak
                \begin{minipage}[t]{6cm}\sloppy} 
                \newcommand\exb{\end{minipage}\kern
                1cm\begin{minipage}[t]{8cm}\sloppy }
                \newcommand\exc{\end{minipage}\kern -3cm
                \smallskip\end{flushleft}}
 
\begin{document} 
 
\begin{center} 
  \Huge 
Global Foliations  \\ of  Vacuum Spacetimes  \\ with $T^2$
  Isometry \normalsize
\end{center} 
 
\exa $\!\!$Beverly K. Berger \\ Department of Physics \\ Oakland
University \\ Rochester, MI 48309 \\ USA \\ 
berger@vela.acs.oakland.edu \\ 
 
\bigskip
 
$\!\!$Piotr T. Chru\'sciel 
\\ D\'epartement de
Math\'ematiques \\  
Facult\'e des Sciences \\ Parc de Grandmont \\ F 37200 Tours \\ France
\\ chrusciel@univ-tours.fr \exb
 
$\!\!$James Isenberg$^1$ \\ Department of Mathematics and \\ Institute
for Theoretical Science \\ University of Oregon \\ Eugene, OR 97403 \\ 
USA \\ jim@newton.uoregon.edu \\ 
 
$\!\!$Vincent Moncrief \\ Department of Physics \\ Yale University \\ 
PO Box 208120 \\ New Haven, CT 06520-8120 \\ USA \\ 
moncrief@hepvms.physics.yale.edu \exc

\vspace{1,5cm}

\abstract{We prove a global existence theorem (with respect to a
  geometrically- defined time) for globally hyperbolic solutions of
  the vacuum Einstein equations which admit a $T^2$ isometry group
  with two-dimensional spacelike orbits, acting on
  $T^3$ spacelike surfaces.
\footnotetext[1]{Visiting
    Scientist: Max-Planck-Institut f\"ur Gravitationsphysik
    (Albert-Einstein-Institut) Schlaatzweg 1, 14473 Potsdam, Germany}
}
 
\newpage
 
\section{Introduction} 
 
One of the ways in which to begin to study the behavior of solutions
of complicated partial differential equation systems like the Einstein
equations is to focus on families of solutions with some prescribed
symmetry. This has long been a practice in general relativity, and
hence much has been learned about solutions of the Einstein equations
which are spherically symmetric \cite{4}, which are spatially
homogeneous \cite{10,19}, which are axisymmetric and stationary
\cite{20}, or have various other prescribed isometries \cite{6,13}.
 
One of the more extensively studied
families of solutions are the Gowdy
spacetimes \cite{11}. Long time existence (in an appropriate sense)
has been proven for them \cite{15}, strong cosmic censorship has been
proven for Gowdy spacetimes which are polarized \cite{9}, and much is
known about the small (but infinite dimensional) set of Gowdy
spacetimes which admit extension across a Cauchy horizon \cite{16,8}.

The Gowdy spacetimes are characterized primarily by their admittance
of a spatially-acting $T^2$-isometry group, but certain other
restrictions are imposed as well (see Section 2) \cite {11}. In this
paper, we examine spacetimes which retain the spatial $T^2$-isometry,
but do not satisfy these other restrictions \cite{5}. Our main result
is that, for this wider class of spacetimes, long-time existence (in a
sense similar to that proven for Gowdy $T^3$ spacetimes) holds.
 
While the notion of global long-time existence is unambiguous for
partial differential equations on a flat background spacetime with a
fixed time measure, this is not true for Einstein's field equations
and general relativity. A solution which remains regular for an
infinite range of one time scale may become singular within a finite
range of another. One way around this problem is to find a choice of
time which is fixed to the geometry of the spacetime. For Gowdy
spacetimes, as well as for the more general $T^2$-symmetric spacetimes
which we study here, there is such a geometric time choice, 
defined (up to proportionality constant) to be
the area of the orbits of the $T^2$ isometry group. If we label this
choice of time $R$, 
then the long-time existence theorem for $T^3$
Gowdy spacetimes \cite{15} says that the maximal spacetime development
of any initial data satisfying certain regularity assumptions admits
the time choice $R$, with $R$ extending from zero to infinity, unless
the spacetime has a $T^3$ isometry group ({\em i.e.}, Kasner), in which case
$R$ may extend only from some positive $R_0$ to $R=\infty$.
 
What we show here is that for the more general $T^2$-symmetric
spacetimes (for which the field equations are significantly more
nonlinear than in the Gowdy case) a similar result is true: There is
always a global $R=t$ foliation, with $R\in (R_0,\infty )$ for some
$R_0\geq 0$. The result, moreover, is sharp.
 
The bulk of this paper is devoted to the proof of this result, which
we state as Theorem 1 in Section 3. Before stating this theorem, we
define (in Section 2) our family of spacetimes and write the general
parametrization for the metrics as well as the corresponding
expressions for the field equations in two different coordinate
choices which we find useful. We divide our discussion of the proof
into three parts. In the first part (carried out in Section 4), we
focus on the ``contracting ($R\searrow 0)$ direction". We establish
the necessary estimates and prove a global existence result for this
direction in terms of an auxiliary choice of time -- ``conformal
coordinate" time. In the second part (Section 5), we prove a number of
geometric results which relate $R$ to the maximal globally hyperbolic
region of a spacetime and to the conformal coordinate representation.
Using these results, we show that we have an $R=t$ foliation covering
the contracting direction. Then in the third part (Section 6), we
focus on the ``expanding \mbox{($R\to\infty )$} direction".  Working
directly with $R=t$ coordinates, we establish some estimates and then
combine these with results from Section 4 to prove the existence of a
global $R=t$ foliation in the expanding direction. This completes the
proof of our main result.  We conclude in Section 7 with a discussion
of questions concerning these spacetimes which we plan to explore
further.

\section{Description and Field Equations for the $T^2$-Symmetric Spacetimes} 
 
The Gowdy spacetimes all admit a $T^2$ isometry group with spacelike
orbits.  The additional condition which characterizes the Gowdy
spacetimes is that the ``twists" associated to the $T^2$ isometry
group must vanish. More specifically, the twist quantities
$$ K_{\!\!\!\!\!\! _{_{(X)}}}:=\epsilon _{\mu\nu\rho\lambda}X^\mu
Y^\nu \nabla^\rho X^\lambda \quad\rm{and}\quad K_{\!\!\!\!\!\!
  _{_{(Y)}}}:=\epsilon _{\mu\nu\rho\lambda}X^\mu Y^\nu \nabla^\rho
Y^\lambda , \eqno(2.1)
$$ where $X$ and $Y$ are any pair of Killing vector fields generating
the $T^2$ group action, must both be zero (as a defining
characteristic) in a Gowdy spacetime.

In this paper we wish to analyze spacetimes in which this condition
on the twists is relaxed: we shall assume that at least one of the quantities
$K_{\!\!\!\!\!\! _{_{(X)}}}$ or $K_{\!\!\!\!\!\!_{_{(Y)}}}$ is non
zero. It is interesting and important to note that in any spacetime
with a $T^2$ isometry, one can \em always \rm replace $X$ and $Y$ by
constant linear combinations of themselves, $\hat X$ and $\hat Y$, and
thereby cause one or the other of the twist quantities
$K_{\!\!\!\!\!\! _{_{(\hat X)}}}$ or $K_{\!\!\!\!\!\! _{_{(\hat Y)}}}$
to be zero. The vanishing of \em both \rm twist quantities is however
independent of such mixing of Killing vector fields, so the Gowdy
spacetimes are unambiguously distinguished. It is also useful to note
that while the Gowdy spacetimes are compatible with
$S^3\times{\cal R}^1$, $S^2\times S^1\times{\cal R}^1,$ as well as
$T^3\times{\cal R}^1$ spacetime manifolds, if one of the twists is
non-vanishing, then the manifold must be $T^3\times{\cal R}^1$.
Hence we shall restrict our attention to $T^2$ symmetric spacetimes on
$T^3\times{\cal R}^1$.
 
There are two choices of coordinates we will find useful for
specifying the form of the metric and the form of the field equations
for these spacetimes. Both sets of coordinates are chosen to be
compatible with the Killing vector fields in the sense that
$X=a\frac{\partial}{\partial x}+b\frac{\partial}{\partial y}$ and $
Y=c\frac{\partial}{\partial x}+d\frac{\partial}{\partial y}$, for a
set of constants $a,b,c,d$ with det$\left( \begin{array}{cc} a & b \\c
    & d \end{array} \right)\neq 0$, and both use $\theta\in S^1$ to
label the remaining spatial coordinate. The two sets differ mainly in
the choice of time slicing. In the first set, which we call the
``areal coordinate system", we set $t=R$, where $R$ is proportional to
the geometric area function of the orbits of the isometry group. In
terms of these coordinates $(x,y,\theta,t)$, we have the following
general form for the metric and for the vacuum field equations.

\bigskip \bf Areal Coordinates \rm

\bigskip \it Metric \rm
 
$$ g=e^{2(\nu-U)}(-\alpha dt^2+d\theta^2) +\lambda e^{2U}[dx
+Ady+(G_{\!\!\!\!\!_{_1}}+AG_{\!\!\!\!\!_{_2}})d\theta
+(M_{\!\!\!\!\!_{_1}}+AM_{\!\!\!\!\!_{_2}}\enspace\! )dt]^2
$$
$$ +\lambda e^{-2U}t^2[dy+G_{\!\!\!\!\!
  _{_2}}d\theta+M_{\!\!\!\!\!_{_2}}\enspace\! dt]^2 \ .\eqno(2.3)
$$ (Here $U,A,\nu,\alpha,
G_{\!\!\!\!\!_{_1}},G_{\!\!\!\!\!_{_2}},
M_{\!\!\!\!\!_{_1}}$ and
$M_{\!\!\!\!\!_{_2}}$ are functions of $\theta\in 
S^1$ and $t\in{\cal R}^+$, 
and $\lambda$ is a strictly positive constant.)

\bigskip \it Einstein Constraint Equations \nopagebreak\rm

\nopagebreak

$$ \nu_t = t[(U_t^2+\alpha
U_\theta^2)+\frac{e^{4U}}{4t^2}(A_t^2+\alpha A_\theta^2)]+
\lambda \frac{e^{2\nu}}{4t^3}\alpha K^2\ , \eqno(2.4a)
$$
$$ \nu_\theta = t[2U_tU_\theta +\frac{e^{4U}}{2t^2}A_tA_\theta]-
\frac{\alpha_\theta}{2\alpha}\ , \eqno(2.4b)
$$
$$ \alpha_t=-\lambda \frac{e^{2\nu}}{t^3}(\alpha K)^2\ , \eqno(2.4c)
$$
$$ K_\theta=0\ , \eqno(2.4d)
$$
$$ K_t=0 \ .\eqno(2.4e)
$$

\it Einstein Evolution Equations \rm \nopagebreak

$$ U_{tt}-\alpha U_{\theta\theta}=-\frac{U_t}{t}+\frac{\alpha_\theta
  U_\theta}{2}+\frac{\alpha_t U_t}{2\alpha}+\frac{e^{4U}}{2t^2}
(A_t^2- \alpha A_\theta^2)\ , \eqno(2.5a)
$$
$$ A_{tt}-\alpha A_{\theta\theta}=\frac{A_t}{t}+\frac{\alpha_\theta
  A_\theta}{2}+\frac{\alpha_t A_t}{2\alpha}-4A_tU_t+4\alpha A_\theta
U_\theta \ ,\eqno(2.5b)
$$
$$ \nu_{tt}-\alpha
\nu_{\theta\theta} = \frac{\alpha_\theta\nu_\theta}{2}+
\frac{\alpha_t\nu_t}{2 \alpha}-\frac{\alpha_\theta^2}{4\alpha}
+\frac{\alpha_{\theta \theta}}{2}
-U_t^2+\alpha U_\theta^2 
$$
$$
+\frac{e^{4U}}{4t^2}(A_t^2-\alpha A_\theta^2)
-\lambda \frac{3e^{2\nu}\alpha}{4t^4} K^2
\ .
\eqno(2.5c)
$$
 
{\em Auxiliary Equations}
 
$$ {M_{\!\!\!\!\!_{_1}}}_{\theta}=
{G_{\!\!\!\!\!_{_1}}}_{t}+A({G_{\!\!\!\!\!_{_2}}}_{t}-
{M_{\!\!\!\!\!_{_2}}}_{\theta}) \ ,\eqno(2.6a)
$$
$$ {M_{\!\!\!\!\!_{_2}}}_{\theta}= {G_{\!\!\!\!\!_{_2}}}_{t}-
\frac{e^{2\nu}}{t^3}\alpha^{1/2}K \ .\eqno(2.6b)
$$ (Note that here and below, we use $t$ and $\theta$ as subscripts on
$U,A,\nu,\alpha$ etc. exclusively to indicate partial derivatives of
these functions. Note also that, without loss of generality, we have
set the twist quantity $K_{\!\!\!\!\!\!_{_{(X)}}}$ equal to zero, and
we have used ``$K$" to label the remaining twist quantity
$K_{\!\!\!\!\!\!_{_{(Y)}}}$.)
 
The other coordinate system we use here chooses $t$ and $\theta$ so
that, if we forget the ``shift"-type metric components $
G_{\!\!\!\!\!_{_1}}, G_{\!\!\!\!\!_{_2}}, M_{\!\!\!\!\!_{_1}} $ and $
M_{\!\!\!\!\!_{_2}}$, then the induced Lorentz metric on the
($\theta,t)$-labeled space of orbits of the $T^2$ isometry group is
conformally flat. Doing this allows us to remove the function
$\alpha(\theta,t)$ from the metric, but requires that we let the
orbital area $R$ be a function of $\theta$ and $t$. So in this system
of coordinates, which we call "conformal coordinates", the metric and
the field equations take the following form.
 
\bigskip  \bf Conformal Coordinates \rm
 
\bigskip \it Metric \rm
$$ g=e^{2(\nu-U)}(-dt^2+d\theta^2) +\lambda e^{2U}[dx
+Ady+(G_{\!\!\!\!\!_{_1}}+AG_{\!\!\!\!\!_{_2}}\enspace\! )d\theta
+(M_{\!\!\!\!\!_{_1}}+AM_{\!\!\!\!\!_{_2}}\enspace\! )dt]^2
$$
$$ +\lambda R^2 e^{-2U}[dy+G_{\!\!\!\!\!
  _{_2}}d\theta+M_{\!\!\!\!\!_{_2}}\enspace\! dt]^2 \ .\eqno(2.7)
$$ (Here, as before, $U,A,\nu,R,
M_{\!\!\!\!\!_{_1}},G_{\!\!\!\!\!_{_1}},G_{\!\!\!\!\!_{_2}}$ 
and $ M_{\!\!\!\!\!_{_2}}$ 
are functions of $\theta$ and $t$.)

\bigskip \it Einstein Constraint Equations \rm
$$ 0=U_t^2+U_\theta^2+\frac{e^{4U}}{4R^2}(A_t^2+A_\theta^2)+
\frac{R_{\theta\theta}}{R}-\frac{\nu_t R_t}{R}-\frac{\nu_\theta
  R_\theta}{R}+\lambda \frac{e^{2\nu}}{4R^4}K^2 \ ,\eqno(2.8a)
$$
$$ 0=2U_t U_\theta +\frac{e^{4U}}{2R^2}A_t A_\theta + \frac{R_{\theta
    t}}{R}- \frac{\nu_\theta R_t}{R}-\frac{\nu_t R_\theta}{R}\ ,
\eqno(2.8b)
$$
$$ K_\theta=0 \ ,\eqno(2.8c)
$$
$$ K_t=0 \ .\eqno(2.8d)
$$
 
{\it Einstein Evolution Equations}
 
$$ U_{tt}-U_{\theta\theta}=\frac{R_\theta U_\theta}{R}-\frac{R_t
  U_t}{R}+ \frac{e^{4U}}{2R^2}(A_t^2-A_\theta^2) \ ,\eqno(2.9a)
$$
$$ A_{tt}-A_{\theta\theta}=\frac{R_t A_t}{R}-\frac{R_\theta
  A_\theta}{R} +4(A_\theta U_\theta - A_t U_t) \ ,\eqno(2.9b)
$$
$$ R_{tt}-R_{\theta\theta}=\lambda \frac{e^{2\nu}}{2R^3}K^2 \ ,\eqno(2.9c)
$$
$$
\nu_{tt}-\nu_{\theta\theta}=U_\theta^2-U_t^2+\frac{e^{4U}}{4R^2}(A_t^2
-A_\theta^2)-\lambda \frac{3e^{2\nu}}{4R^4}K^2\ . \eqno(2.9d)
$$
 
{\em Auxiliary Equations}
$$ {M_{\!\!\!\!\!_{_1}}}_\theta=
{G_{\!\!\!\!\!_{_1}}}_t+A({G_{\!\!\!\!\!_{_2}}}_t -
{M_{\!\!\!\!\!_{_2}}}_\theta)\ , \eqno(2.10a)
$$
$$ {M_{\!\!\!\!\!_{_2}}}_\theta=
{G_{\!\!\!\!\!_{_2}}}_t-\frac{e^{2\nu}}{R^3}K \ .\eqno(2.10b)
$$ (Our conventions regarding subscripts, and our handling of the
twist quantities, are the same here as above.
 
It has been proven in earlier work \cite{5} that, at least locally,
any globally hyperbolic $T^2$-symmetric spacetime on
$T^3\times{\cal R}^1$ admits each of these coordinate forms (unless
the spacetime is flat). Since our notion of long time existence is
tied to the orbital area function $R$, the statement of our main
result focuses on $R$ and the areal coordinate system (with $R=t$).
Indeed it establishes that areal coordinates cover any globally
hyperbolic $T^2$- symmetric spacetime. To prove this in the expanding
direction, areal coordinates are used directly. However to prove this
in the contracting direction, we find that the conformal coordinate
form is the most useful for carrying out the analysis.
 
\section{The Long Time Existence Theorem} 
 
Let us call ($\gamma,\pi$) \em $T^2$-symmetric initial data on $T^3$
\rm if a) $\gamma$ is a Riemannian metric on $T^3$, invariant under an
effective $T^2$ action; b) $\pi$ is a symmetric 2-tensor on $T^3$,
also invariant under the same $T^2$ group action; and c)
$(\gamma,\pi)$ together satisfy the Einstein constraint equations.
 
To avoid unnecessary details, we will assume that $(\gamma,\pi)$ are
smooth ($C^\infty)$ on $T^3$. Our result holds for weaker
differentiability conditions imposed on $(\gamma,\pi)$, but we will
not state those conditions here. Our main result is the following
 
\bigskip \bf Theorem 1 \rm
Let ($\gamma,\pi)$ be a set of smooth $T^2$ symmetric initial data on
$T^3$.  For some non-negative constant $c$, there exists a globally
hyperbolic spacetime ($M^4,g)$ such that \\ (i)
$M^4=T^3\times(c,\infty)$, \\ (ii) $g$ satisfies the vacuum Einstein
equations, \\ (iii) $M^4$ is covered by areal coordinates
($x,y,\theta,t)$, with $t\in (c,\infty)$, so the metric globally takes
the form (2.3), \\ (iv) $(M^4,g)$ is isometrically diffeomorphic to the
maximal globally hyperbolic development of the initial data
($\gamma,\pi)$.
 
\bigskip \bf Outline of Proof: \rm  
As noted in the introduction, we carry out our proof of Theorem 1 in
Sections 4, 5 and 6. The logic of the proof is as follows: First (in
Section 4), working in terms of conformal coordinates, we look at the
evolution of a solution from the data ($\gamma,\pi$) toward the
contracting direction. (Without loss of generality, we may choose the
time-orientation so that this is towards the past.) Using the field
equations and light cone arguments to establish a number of estimates
for various components of the metric, we go on to prove (using results
from, e.g. \cite{14}) a global existence result in the following
sense: So long as $R$ stays bounded away from zero, the past maximal
development of ($\gamma,\pi$) in terms of conformal coordinates -- let
us label it $D_{conf}^-(\gamma,\pi)$ -- has $t\to -\infty$.  The
evolution (in terms of conformal coordinates) stops only if $R$
approaches zero.
 
Next (in Section 5) we prove a number of geometric results, most of
which concern the behavior of the orbital area function $R$ in a
globally hyperbolic spacetime such as $D_{conf}^- (\gamma,\pi)$. The
first of these, a Killing vector argument, shows that $R$ is positive
everywhere in the globally hyperbolic region of a $T^2$ symmetric
spacetime. The next one shows that along any past inextendible
timelike path in $D_{conf}^- (\gamma,\pi)$, $R$ approaches a limit
$R_0 \geq 0 $ (to be identified with ``$c$" in \mbox{Theorem 1}).
Moreover, one has the same limit along all such paths. Combining this
result with another (proven in Section 5) which shows that for any
$\rho\in(R_0,R_1)$ -- where $R_1$ is the minimum value of $R$ on the
initial surface with data $(\gamma,\pi)$ -- the $R=\rho$ level set in
$D_{conf}^- (\gamma,\pi)$ is a Cauchy surface, we can argue that $
D_{conf}^- (\gamma,\pi )$ admits areal coordinates, at least to the
past of the hypersurface with constant $R=R_1$. To show that $
D_{conf}^-(\gamma,\pi)$ is isometrically diffeomorphic to the maximal
\cite{2} globally hyperbolic past development $ D^-(\gamma,\pi)$ of
$(\gamma,\pi)$ on $T^3$, we need two further geometric results. We
show that if $t\to -\infty$ in $ D_{conf}^-(\gamma,\pi)$, then $
D_{conf}^-(\gamma,\pi) \approx D^-(\gamma,\pi)$, and we show that if
$R\to 0$ in $ D_{conf}^-(\gamma,\pi)$, then $
D_{conf}^-(\gamma,\pi)\approx D^-(\gamma,\pi)$. This completes the
argument that to the past of the hypersurface with $R=R_1$, $
D^-(\gamma,\pi)$ can be covered by areal coordinates. Note, that we
also prove in Section 5 that if $R\to\infty$ in any future development
of $(\gamma,\pi)$, then that future development is maximal.
 
Our proof is more direct for the expanding, future direction. Based on
results from Section 5, we have an $R=$ constant Cauchy surface -- say
$R=R_2$ -- to the past of our original Cauchy surface. Let us call the
data on this surface $(\gamma_1,\pi_1)$. Then working in areal
coordinates, we can use the field equations and a light cone arguments
to establish a collection of estimates, and from these (and, again,
results from \cite{14}) we prove a global existence result which says
that the future maximal development of $(\gamma_1,\pi_1)$ in terms of
areal coordinates -- $D_{areal}^+(\gamma,\pi)$ -- has $t=R\to\infty$.
As shown in Section 5, it follows that $
D_{areal}^+(\gamma_1,\pi_1)\approx D^+(\gamma_1,\pi_1)$. This
completes the proof of Theorem 1.
 
Note that since Theorem 1 is known to be true in the Gowdy case
\cite{15} ({\em cf.} also \cite{5}), we will henceforth presume that
the twist quantity $K$ is 
\em not \rm zero. Note also that in both areal and conformal
coordinates, the constraint equations require that $K$  be constant
on spacetimes.
 
\section{Analysis in the Contracting Direction} 
 
As noted above in the outline of the proof of Theorem 1, our goal in
this section is to show that, so long as the orbit area function $R$
stays bounded away from zero, the past (contracting direction)
development of $(\gamma,\pi)$ in terms of conformal coordinates -- the
spacetime region we call $D_{conf}^-(\gamma,\pi)$ -- has $t\to-\infty$.
Our argument for this begins by recalling the local existence result
for $T^2$-symmetric solutions of the Einstein equations in conformal
coordinate form. This result (see Lemma 4.2 in \cite{5}) tells us that
for any $T^2$-symmetric initial data $(\gamma,\pi)$ on $T^3$, we can
always find an interval $(t_1,t_2)$ and real valued $C^\infty$
functions $ R,U,A,\nu, G_{\!\!\!\!\!_{_1}}, G_{\!\!\!\!\!_{_2}},
M_{\!\!\!\!\!_{_1}} $ and $ M_{\!\!\!\!\!_{_2}}$ on $T^3
\times (t_1,t_2)$ such that (1) these functions satisfy the
Einstein equations (2.8)-(2.10) in conformal coordinate form; and (2)
for some $t_0\in(t_1,t_2)$, the spacetime metric $g$ constructed from
$(R,U,A,\nu, G_{\!\!\!\!\!_{_1}}, G_{\!\!\!\!\!_{_2}},
M_{\!\!\!\!\!_{_1}}, M_{\!\!\!\!\!_{_1}}\enspace\! )$ according to
equation (2.7) induces initial data on the $t_0- $slice which is
smoothly spatially diffeomorphic to $(\gamma,\pi)$.
 
With this established, it follows as a consequence of standard
longtime existence theorems from PDE theory (see, e.g., Theorems 2.1
and 2.2, and Corollaries 1 and 2 in Chapter 2 of \cite{14}), 
that to
show that these fields extend to $t\to - \infty$ as a solution of
(2.8)-(2.10), it is sufficient to verify the following: For any finite
interval $(\tau,t_0)$ on which they exist as a solution to
(2.8)-(2.10), the functions $(R,U,A,\nu, G_{\!\!\!\!\!_{_1}},
G_{\!\!\!\!\!_{_2}}, M_{\!\!\!\!\!_{_1}},
M_{\!\!\!\!\!_{_1}}\enspace\! )$ and their first and second
derivatives are uniformly bounded.
 
So that is our task here: to establish these $C^2$ bounds. We do this
in a series of steps.
 
\bigskip \underline{Step 1} (Monotonicity of $R$ and Bounds on its
First Derivatives)
 
A key first step here is the verification that the function $R$ and
its first derivatives are controlled. The argument for this control
starts with Theorem 4.1 of \cite{5}, where it is shown that it follows
from the constraint equations in conformal coordinate form (2.8) that
$\nabla R$ is timelike (i.e., $g(\nabla R,\nabla R)<0$). Note that if
$\nabla R$ is timelike everywhere, one must have $R_t$ nonzero
everywhere. One could have $R_t>0$ or $R_t<0$, but our choice of
time-orientation -- the past corresponds to contracting $T^2$ orbits --
dictates $R_t > 0$.  Thus along any past directed causal path, $R$ monotonically
decreases.
 
To show that the first derivatives of $R$ are bounded along any such
path, one uses the wave equation (2.9c) for $R$, which takes the form
$$ \partial_\lambda R_\xi =\lambda \frac{e^{2\nu}}{4R^3}K^2, \eqno(4.1a)
$$ or equivalently
$$ \partial_\xi R_\lambda =\lambda \frac{e^{2\nu}}{4R^3}K^2, \eqno(4.1b)
$$ in terms of null\footnote{These are null coordinates for the
  two-dimensional base spacetime ($S^1\times{\cal R}^+,{}^2g)$ with
  the conformally flat metric $g=e^{2(\nu-U)}(-dt^2+d\theta^2)$.}
coordinates
$$ \lambda=\frac{1}{\sqrt{2}}(t+\theta)\ , \qquad\qquad
\xi=\frac{1}{\sqrt{2}}(t- \theta).  \eqno(4.2)
$$ Since the right hand side of (4.1a) is positive, it follows that if
we start at any point $(\theta_0,t_0)$ on the initial surface and
follow back along the path $(\theta_0+s,t_0-s)$ generated by
$\partial_\lambda$, then for any $s>0$, we have
$$ R_\xi(\theta_0+s,t_0-s)<R_\xi(\theta_0,t_0).  \eqno(4.3a)
$$ Similarly, we obtain (from 4.1b)
$$ R_\lambda(\theta-s,t_0-s)<R_\lambda(\theta_0,t_0)\ . \eqno(4.3b)
$$ While it does not follow that $R_\xi$ and $R_\lambda$ decrease with
decreasing $t$ for fixed $\theta_0$, it \em does \rm follow from (4.3)
that $R_\xi$ and $R_\lambda$ are bounded into the past in the
following sense:
$$ R_\xi(\hat\theta,\hat t)<\max_{\theta\in S^1} R_\xi(\theta,t_0)\ ,
\qquad R_\lambda(\hat\theta,\hat t)<\max_{\theta\in S^1}
R_\lambda(\theta,t_0)\ . \eqno(4.4)
$$ for any $\hat t<t_0$, and for any $\hat\theta$. Since
$R_t=\frac{1}{\sqrt{2}}(R_\xi+R_\lambda)$, we therefore have
$$ R_t(\hat\theta,\hat t)<\max_{\theta\in S^1} 
\left( R_\xi+R_\lambda\right)(\theta,t_0)  \ ,
\eqno(4.5)
$$ 
for any $\hat t<t_0$, and for any $\hat\theta$.
 
There is no equivalent result for $R_\theta$. But since $\nabla R$ is
timelike, we have $\mid R_\theta\mid <\mid R_t\mid$ everywhere. Thus
we find that $\mid R_\theta\mid$, as well as $R_t$, is bounded into the
past, and we conclude that $R$ is uniformly $C^1$-bounded to the past
of the initial data surface.
 
\bigskip \underline{Step 2} (Bounds on $U,U_\theta ,T_t ,$ and $A,
A_\theta ,A_t$.)
 
The method we use to argue that quantities like $U$ and $A$ -- which
satisfy nonlinear coupled wave-type evolution equations -- are
controlled to the past of the initial data surface for all
$t\in(\tau,t_0)$, is called the ``light cone estimate" method
\cite{15}. The basic idea of this method is to first show that the
evolution equations for the quantities of interest -- say $U$ and $A$
-- imply that $U$ and $A$ satisfy equations of the form
$$ n(E+P)=J \ ,\eqno(4.6a)
$$
$$ l(E-P)=L\ , \eqno(4.6b)
$$ where $E$ and $P$ are quadratic functions in the first derivatives
of $U$ and $A$, where $J$ and $L$ are composed of functions which are
bounded on the region of interest, and possibly of first derivatives
of $U$ and $A$ as well, and where $n$ and $l$ are a pair of
independent null tangent vectors. One then formally integrates these
equations (4.6) along null paths generated by $n$ and $l$; using
Gromwall's lemma as in \cite{5}, 
one thereby derives estimates for the first
derivatives of $U$ and $A$ for any value of $t\in(\tau,t_0)$, in terms
of their values at $t_0$.
 
To obtain equations of the form (4.6) for $U$ and $A$, it is useful to
first rewrite their evolution equations in terms of a \em wave map.
\rm
So we consider a base Lorentzian manifold (${}^2M_\tau,{}^2\eta)$
with the two-dimensional manifold ${}^2M_\tau$ corresponding to the
past conformal coordinate development of $(\gamma,\pi$) -- 
with coordinates $(\theta,t)$  in  $S^1\times {{\cal R} }$,
or a subset thereof
-- 
and with the metric
$$ {}^2\eta:=-dt^2+d\theta^2 ; \eqno(4.7a)
$$ and we consider a family of target Riemannian manifolds
$({\cal R},h_{(\theta,t)})$ with
$$ h_{(\theta,t)}:=R(\theta,t)dU^2+\frac{e^{4U}}{4R(\theta,t)}dA^2.
\eqno(4.7b)
$$ (Note the explicit ($\theta,t)$ dependence of the target manifold
metric; this makes our set-up slightly different from a standard wave
map, but the difference is easily handled). 
The maps we consider take the form
\begin{eqnarray}
 \Phi :\  {}^2 M  \ & \longrightarrow & {\cal R}^2\ ,
\nonumber \\
\setcounter{equation}{8}
 (\theta,t) & \longmapsto & \Phi(\theta,t)=\left({U(\theta,t)} 
  {A(\theta,t)}\right) \ , 
\end{eqnarray}
with $ h_{(\theta,t)}$ providing an inner product on their
tangents; e.g.,
$$ <\Phi_\theta ,\Phi_t>=h_{ab}\Phi^a_\theta \Phi^b_t= RU_\theta U_t
+\frac{e^{4U}}{4R}A_\theta A_t.  \eqno(4.9)
$$
 There is a covariant derivative $D$ compatible with ${}^2\eta$ 
and semi-compatible with $h(\theta,t)$. Using Greek indices for the base
($\mu\leftrightarrow\theta,t)$ and Latin indices for the target
($a\leftrightarrow U,A)$, we express the action of $D$ as follows:
$$ D_\nu\Phi_\mu^a=\partial_\nu\Phi_\mu^a + \Gamma^a_{\enspace bc}
\Phi^b_\mu\Phi^c_\nu - \Gamma^\lambda_{\enspace\mu\nu}\Phi^a_\lambda
\ ,
\eqno(4.10)
$$ with the base Christoffel coefficients
$\Gamma^\lambda_{\enspace\mu\nu}$ vanishing (in Section 6, we will
work with a wave map for which these are not zero), and with the target
Christoffel coefficients taking the values
$$ \Gamma^U_{\enspace UU}=0\ ,\quad \Gamma^U_{\enspace UA}=0 \ ,\quad
\Gamma^U_{\enspace AA}=-\frac{e^{4U}}{2R^2}\ ,
$$
$$ \Gamma^A_{\enspace AA}=0\ ,\quad \Gamma^A_{\enspace AU}=2 \ ,\quad
\Gamma^A_{\enspace UU}=0 \ .\eqno(4.11)
$$ As noted, $D$ is compatible with the flat metric ${}^2g$, but not
with $h_{(\theta,t)}$ because of the explicit $\theta$ and $t$
dependence (through $R$); we have
$$ D_\mu h_{ab}=R_\mu (\delta_a^U \delta_b^U -
\frac{e^{4U}}{4R^2}\delta_a^A\delta_b^A )\ .  \eqno(4.12)
$$

The covariant derivative $D$ defines a wave operator $\Box
:=g^{\mu\nu} D_\mu D_\nu$ on our maps. Using this operator, the
evolution equations (2.9a)-(2.9b) for $U$ and $A$ take the form
$$ \Box U=\frac{U_t R_t}{R}-\frac{U_\theta R_\theta}{R}\quad ,
\quad \Box
A=-\frac{A_t R_t}{R}+\frac{A_\theta R_\theta}{R} \eqno(4.13)
$$ which we may write jointly as
$$ \Box\Phi^a=\psi^a \eqno(4.14a)
$$ where
$$ \psi^a=\left({\frac{U_t R_t}{R}-\frac{U_\theta R_\theta}{R}}
  \above0pt {-\frac{A_t R_t}{R}+\frac{A_\theta R_\theta}{R}} \right).
\eqno(4.14b)
$$
 
To derive equations of the form (4.6) from the wave equations (4.14),
we now define an ``energy-momentum-tensor" for the maps $\Phi$:
\begin{eqnarray*} 
  T_{\mu\nu}: & = &
  <\Phi_\mu,\Phi_\nu>-\frac{1}{2}g_{\mu\nu}g^{\alpha\beta}
  <\Phi_\alpha,\Phi_\beta> \\ & = & RU_\mu
  U_\nu+\frac{e^{4U}}{4R}A_\mu A_\nu + \frac{1}{2}
  g_{\mu\nu}[R(U_t^2-U_\theta^2)+\frac{e^{4U}}{4R}(A_t^2-A_\theta^2])
\ .
\end{eqnarray*}
$$ \eqno(4.15)
$$ One notes the components of $T_{\mu\nu}$
$$ T_{tt}=\frac{1}{2}R(U_t^2+U_\theta^2)+\frac{e^{4U}}{8R}
(A_t^2+A_\theta^2)\ , \eqno(4.16a)
$$
$$ T_{\theta t}=RU_t U_\theta +\frac{e^{4U}}{4R}A_t A_\theta
\ ,\eqno(4.16b)
$$ and
$$ T_{\theta\theta}=T_{tt}\ ; \eqno(4.16c)
$$ $T_{tt}$ will be our quantity $E$ in (4.6) while $T_{t\theta}$ will
be $P$. One also notes the formula for the covariant divergence of
$T_{\mu\nu}$
$$ D_\nu T_\mu^{\enspace\nu}=<\partial_\mu\Phi,\psi>+U_\mu U^\nu
(\partial_\nu R)-\frac{1}{2}U^\alpha U_\alpha (\partial_\mu R)
$$
$$ +\frac{e^{4U}}{4R}(\frac{1}{2}A^\alpha A_\alpha(\partial_\mu
R)-A_\mu A^\nu(\partial_\nu R)) \ ,\eqno(4.17)
$$ where $\psi$ is defined in (4.14b), and where all but the first
term on the right hand side of (4.17) appear because of the $\theta$
and $t$ dependence of $h_{(\theta ,t)}$ (see equation (4.12)).
 
Since the base metric ${}^2\eta$ is flat, one readily identifies null
vectors and the corresponding null coordinates: We use $\lambda$ and
$\xi$ as in (4.2) for the coordinates, and $n=\partial_\xi$ and
$l=\partial_\lambda$ for the corresponding vectors. It is now
straightforward to show that
$$ D_\nu T^\nu_{\enspace n}=-D_lT_{nn}\ ,\qquad D_\nu T^\nu_{\enspace
  l}=-D_n T_{ll}\ . \eqno(4.18)
$$ Then combining (4.17) and (4.18), we obtain the formulas
$$ \partial_\xi T_{\lambda\lambda}=n(T_{\lambda\lambda})=
-\frac{R_\lambda}{2\sqrt{2}}[(U_t^2-U_\theta^2)+
\frac{e^{4U}}{4R^2}(-A_t^2+A_\theta^2)] \ ,\eqno(4.19a)
$$ and
$$ \partial_\lambda T_{\xi\xi}=l(T_{\lambda\lambda})=
-\frac{R_\xi}{2\sqrt{2}}[(U_t^2-U_\theta^2)+
\frac{e^{4U}}{4R^2}(-A_t^2+A_\theta^2)] \ .\eqno(4.19b)
$$
 
Now, since we note that $T_{\lambda\lambda}=T_{tt}+T_{t\theta}$ and
$T_{\xi\xi}=T_{tt}-T_{t\theta}$, and since we note that the
expressions on the right hand side of (4.19) involve only terms
quadratic in first derivations of $U$ and $A$, along with quantities
(first derivatives of $R$) which are bounded in ${}^2M$, we see that
if we set $E=T_{tt}$, $P=T_{t\theta}$, and $J=$[the right hand side of
(4.19a)] and $L=$[the right hand side of (4.19b)], then (4.19) matches
(4.6).
 
We now describe how one uses equations (4.19) to obtain estimates for
$E=T_{tt}$ at any $(\hat\theta,\hat t)\in{}^2M_\tau$, in terms of $E$
and other controlled quantities on the initial surface. Such estimates
control the derivatives of $U$ and $A$ in terms of data at $t_0$.
 
We start by formally integrating both equations (4.19) -- in the form
$\partial_\xi(E+P)=J$ and $\partial_\lambda (E-P)=L$ of (4.6) -- along
null paths which start at $(\hat\theta,\hat t)$ and end back on the
initial data surface.  Adding the results of these integrations, we
have
\setcounter{equation}{19}
\begin{eqnarray} 
  E(\hat\theta,\hat t) & = & E(\hat\theta+(t_0-\hat
  t),t_0)+E(\hat\theta -(t_0 -\hat t),t_0)\nonumber \\ & + &
  P(\hat\theta+(t_0-\hat t),t_0) -P(\hat\theta-(t_0-\hat t),t_0)\nonumber \\ &
  + & \int\limits_{t_0}^{\hat t}[\tau(\hat\theta+(s-\hat t),
  s)+L(\hat\theta-(s-\hat t),s)]ds \ .
\end{eqnarray}
We next take supremums\footnote{The supremums over $\theta$ include all values
  for which the solution exists for a given value of $t$. For $t_0$,
  we have $\theta\in S^1$.  For smaller values of $t$, we generally
  have only a portion of $S^1$.} over all values of the space
coordinate $\theta$ on both sides of (4.20). Carefully noting the
explicit forms of $J$ and $L$ (see equations (4.19)), and using
various straightforward inequalities, we obtain
\begin{eqnarray}
  \sup_{\theta}E(\theta,\hat t) & \leq & 2\sup_{\theta}E(\theta,t_0)+
  2\sup_{\theta}P(\theta,t_0)\nonumber \\ & + & \int\limits_{t_0}^{\hat
    t}\beta(s)\sup_{\theta}E(\theta,s)ds\ , 
\end{eqnarray} 
where $\beta(s)$ is a bounded function (related to the bounds of
$R_t$ and $R_\theta$).
 We now apply Gromwall's lemma (see, e.g., Lemma 3.23 in \cite{5}) to
(4.21); we get the inequality
$$ \sup_{\theta}E(\theta,\hat t)\leq[2\sup_{\theta}E(\theta,t_0)+
2\sup_{\theta}P(\theta,t_0)]\exp(\int\limits_{t_0}^{\hat
  t}\beta(s)ds).  \eqno(4.22)
$$ Since $\beta(t)$ and therefore $\exp(\int_{t_0}^{\hat t}\beta(s)ds$
are bounded, (4.22) provides the desired bounds on $\mid U_\theta\mid
,\mid U_t\mid ,\mid \frac{e^{2U}}{2R}A_\theta\mid$ and
$\mid\frac{e^{2U}}{2R}A_t\mid $ for all $t\in (\tau,t_0]$.
 
Once we have bounds on the first derivatives of $U$, it immediately
follows (by integration over appropriate paths in ${}^2M_\tau$) that
$U$ is bounded for $t\in(\tau,t_0]$ as well. Then, so long as $R$ is
bounded away from zero on ${}^2M_\tau$, we obtain bounds on $A_\theta$
and $A_t$ and consequently on $A$. We thus have uniform $C^1$ bounds
on $U$ and $A$ in ${}^2M_\tau$, so long as $R$ is bounded away from
zero.
 
\bigskip \underline{Step 3} (Bounds on $\nu,\nu_\theta,\nu_t$)
 
With the first derivatives of $U$ and $A$ bounded as argued in
\mbox{Step 2}, it appears as if the evolution equation (2.9d) for
$\nu$ might imply that $\nu$ and its first derivatives can be
controlled by applying a light cone estimate. This cannot be done
directly, however, because the last term on the right hand side of
equation (2.9d) -- 
$\lambda \frac{e^{2\nu}}{4R^4}K^2$ --
is not controlled by previous estimates.
 
To handle this, we consider the quantity
$$ \beta:=\nu+ \frac{3}{2} \ln R.  \eqno(4.23)
$$ Using the evolution equations for $\mu$ and $R$, we readily derive
one for $\beta$:
$$ \beta_{tt}-\beta_{\theta\theta}=U_{\theta}^2-U_t^2+
\frac{e^{4U}}{4R^2}(A_t^2-A_\theta^2)-\frac{3}{2R^2}(R_t^2 -R_\theta^2)\ .
\eqno(4.24)
$$ We could apply a light cone estimate argument (as in step 2) to
this equation, but since the right hand side of (4.24) contains
nothing but controlled quantities (and does not contain $\beta$ or its
derivatives), we argue more simply as follows: 
We first use the null coordinates $\xi$ and $\lambda$ to rewrite
(4.24) in the form
$$ \partial_\lambda\beta_\xi=Z\ , \eqno(4.25)
$$ where $Z$ is just the right hand side of (4.24). Choosing an
arbitrary point $(\hat\theta,\hat t)\in{}^2M_\tau$, we obtain
$\beta_\xi(\hat\theta,\hat t)$ by integrating up along the null path
$(\hat\theta -s,\hat t +s)$ which ends at $(\hat\theta+\hat t-t_0,t_0).$
Since $Z$ is bounded in ${}^2M_\tau$ (so long as $R$ is bounded away
from zero), $\beta_\xi$ is bounded in ${}^2M_\tau$. Similarly we argue
from the equation $\partial_\xi\beta_\lambda=Z$ that $\beta_\lambda$
is bounded in ${}^2M_\tau$. Since
$\beta_t=\frac{1}{\sqrt{2}}(\beta_\lambda+\beta_\xi) $ and
$\beta_\theta=\frac{1}{\sqrt{2}}(\beta_\lambda-\beta_\xi)$, it follows
that $\beta_t$ and $\beta_\theta$ are bounded as well. But
$\nu_t=\beta_t- \frac{3}{2}\frac{R_t}{R}$ and
$\nu_\theta=\beta_\theta-\frac{3}{2}\frac{R_\theta}{R}$, so we have
shown that the first derivatives of $\nu$ -- and consequently $\nu$
itself -- must be uniformly bounded for $t\in(\tau,t_0]$. We now have
uniform $C^1$ bounds on all of the principal quantities.
 
\bigskip \underline{Step 4} (Bounds on Second Derivatives of $R$)
 
The constraint equations (2.8a) and (2.8b) allow one to express
$R_{\theta\theta}$ as well as $R_{\theta t}$ algebraically as
functions of $R,U,A$ and $\nu$, and their first derivatives. It
follows immediately that $R_{\theta\theta}$ and $R_{\theta t}$ are
bounded, so long as $R$ is bounded away from zero. It then follows
from the wave equation (2.9c) for $R$ that 
$R_{tt}=R_{\theta\theta}+\lambda \frac{e^{2\nu}}{2R^3}K^2$ is also
bounded, so long as $R$ bounded away from zero.
 
\bigskip \underline{Step 5} (Bounds on Second Derivatives of
$U,A,\nu$)
 
It should be clear from the forms of the evolution equations (2.9a)
and (2.9b) for $U$ and $A$ that if we take time derivatives of both of
these equations we obtain evolution equations for $U_t$ and $A_t$
which are of the appropriate form for applying light cone estimates.
We thereby obtain uniform bounds on $U_{tt},U_{t\theta},A_{tt}$ and
$A_{t\theta}$. Bounds on $U_{\theta\theta}$ and $A_{\theta\theta}$
then follow from (2.9a) and (2.9b) directly.
 
Similarly, if we take the time derivative of equation (2.9c), the
evolution equation for $\nu$, we obtain a wave equation for $\nu_t$ to
which the analysis described in Step 3 can be applied, giving us
bounds on $\nu_{tt}$ and $\nu_{t\theta}$ (Note that for these $C^2$
bounds on $\nu$, we can work directly with $\nu$ rather having to work
with $\beta$). We then bound $\nu_{\theta\theta}$ using equation
(2.9d).
 
We could continue to obtain bounds on higher derivatives of these
quantities (along with $R$); however, to apply the global existence
theorem cited in \cite{14}, $C^2$ bounds are sufficient. Thus we have
shown that so long as $R$ stays bounded away from zero, the functions
$R,U,A$ and $\nu$ extend (as solutions of equations (2.8)-(2.9)) to
$t\to -\infty$.
 
\bigskip \underline{Step 6} (Extension of the Shift Functions)
 
We still need to show that the ``shift" functions $
G_{\!\!\!\!\!_{_1}}, G_{\!\!\!\!\!_{_2}}, M_{\!\!\!\!\!_{_1}}$ and $
M_{\!\!\!\!\!_{_2}}$  extend to 
$D_{conf}^- (\gamma,\pi)$  (which we have defined to be the maximal
subset of $T^3 \times {{\cal R}}$ on which the functions $(R,U,A,\nu)$ 
are solutions 
of (2.8)--(2.9), with the corresponding spacetime metric assuming
initial data $ (\gamma,\pi)$). The only equations 
in which they appear are the auxiliary constraint equations (2.10). 
Recall that in the
Cauchy problem for the 
Einstein equations, the ``shift'' functions have a gauge character,
corresponding to the freedom of propagating the coordinate
system from the initial value hypersurface to the spacetime.
In the gauges presented above, that freedom of propagating the
coordinates $x^a$ still persists. So for an arbitrary choice
throughout $D_{conf}^- (\gamma,\pi)$ of the ``shift'' functions 
$M_{\!\!\!\!\!_{_1}}\enspace (\theta,t)$ and
$M_{\!\!\!\!\!_{_2}}\enspace  (\theta,t)$, and for any initial data
for  $ G_{\!\!\!\!\!_{_1}}$ and $ G_{\!\!\!\!\!_{_2}} $, we can
integrate (2.10) in time to obtain
$$
{G_{\!\!\!\!\!_{_1}}}(\theta,t_1)= 
\int_{t_0}^{t_1} \left[-\frac{Ae^{2\nu}K}{R^3}+
M_{\!\!\!\!\!_{_1}}\enspace_\theta \right](\theta,\tau)d\tau +
{G_{\!\!\!\!\!_{_1}}}(\theta,t_0)
\ , \eqno(4.26a)
$$ 
$$
{G_{\!\!\!\!\!_{_2}}}(\theta,t_1)= 
\int_{t_0}^{t_1} \left[\frac{e^{2\nu}K}{R^3}+
M_{\!\!\!\!\!_{_2}}\enspace_\theta \right](\theta,\tau)d\tau +
{G_{\!\!\!\!\!_{_2}}}(\theta,t_0)
 \  . \eqno(4.26b)
$$ 
We now have a solution of the full set of Einstein equations
(2.8)-(2.10) throughout $D_{conf}^- (\gamma,\pi)$.
We shall show in Section 5 that this set is the maximal past
development of our chosen $T^2$--symmetric initial data.

\section{$\!\!\!$\mbox{Some Geometric Results concerning $R$}} 
 
In Section 4, we have shown that if $R$ stays bounded away from zero,
then $D_{conf}^- (\gamma,\pi)$ has $t\to -\infty$. In this section, we
prove a collection of geometric results which allow us to conclude
that, whether or not $R$ stays bounded away from zero,
$D_{conf}^-(\gamma,\pi)$ admits a foliation by areal coordinates, and
also covers the past maximal globally hyperbolic development
$D^-(\gamma,\pi)$ of $(\gamma,\pi)$.
 
The first of these results concerns zeroes of $R$
  
\bf Proposition 1  \it 
Let (M,g) be a globally hyperbolic development of $T^2$-symmetric
initial data.  The $T^2$ orbital area function $R$ is positive
everywhere in (M,g).  \rm
 
\bigskip \bf Proof: \rm
 The proof of this proposition depends upon two lemmas, which have
wider application than our $T^2$-symmetric case
 
\bigskip \bf Lemma 1.1\rm
 \it Let (M,g) be a globally hyperbolic development of initial data
($\Sigma^3,\gamma,\pi)$. Assume that the initial data is invariant
under the action of a compact group G. Let X be any Killing
vector field on (M,g) which is generated by the action of G on (M,g).
Then wherever X is non vanishing, it is spacelike.
 
\bigskip \bf Proof of Lemma 1.1: \rm
 In \cite{7}, it is shown that indeed the action of the group $G$
extends from the initial data hypersurface to the spacetime. So
($M,g)$ is $G$-invariant, and at least the initial hypersurface is
invariant under the action. We now construct a foliation of $(M,g)$ by
spacelike hypersurfaces which is also $G$-invariant.
 
Let $t$ be any time function on the globally hyperbolic spacetime
$(M,g)$. Then we may define a new function on $M$
$$ \hat t:=\int\limits_G t\cdot\phi_gd\mu_g \eqno(5.1)
$$ where $\phi_g$ denotes the action of $G$ on $M$, and $d\mu_g$ is
the Haar measure. Since the spacetime metric is invariant under $G$,
we verify that $\hat t$ is also a time function, and its level
hypersurfaces $\Sigma_{\hat t}$ are Cauchy surfaces. It follows from
the definition of $\hat t$ that the hypersurfaces $\Sigma_{\hat t}$
are invariant under $G$. Hence, the Killing vector fields generated by
$G$ are tangent to the spacelike hypersurfaces $\Sigma_{\hat t}$.  The
result follows.
\hfill   $\Delta$

\bf Lemma 1.2\rm
 \it Let (M,g) be a globally hyperbolic spacetime, and let
$\phi_\lambda$ be a one- parameter group of isometries of (M,g) which
leaves a particular Cauchy surface $\Sigma_0$ invariant. Let
$X=\frac{d}{d\lambda}\phi_\lambda\mid_{\lambda=0}$ be the Killing
vector field which generates $\phi_\lambda$. If for some point
$p\in\Sigma_0$ one has $X(p)=0$, then on every Cauchy surface
$\hat\Sigma$ in $M$ there exists a point $\hat p\in\hat\Sigma$ such
that $X(\hat p)=0$. \rm

\bigskip \bf Proof of Lemma 1.2: \rm
 Let $\Gamma(s)$ be a maximally extended, affinely parametrized
timelike geodesic such that $\Gamma(0)=p$ and $\frac{d}{ds}\Gamma(0)$
is a unit length vector normal to $\Sigma_0$. Since $\Sigma_0$ is
invariant under $\phi_t$, the unit length vector field $e_\perp(0)$
normal to $\Sigma_0$ is also invariant under $\phi_t$; it follows that
$\phi_\lambda(\Gamma)=\Gamma$.  Since the affine parameter is
invariant under $\phi_\lambda$, we further have
$\phi_\lambda(\Gamma(s))=\Gamma(s)$ for all $s$ and $\lambda$.
 
We now pick a Cauchy surface $\hat \Sigma$. Since $\Gamma(s)$ is an
inextendible timelike path, it must intersect $\hat\Sigma$ at some
point $\hat p=\Gamma(\hat s).$ But
$$ \phi_\lambda(\hat p)=\phi_\lambda(\Gamma(\hat s))=\Gamma(\hat
s)=\hat p \eqno(5.2)
$$ so we have $X(\hat p)=0$.
\hfill 
  $\Delta$

Note that Lemma 1.2 tells us that if a spacetime Killing field tangent
to a particular Cauchy surface in a globally hyperbolic spacetime
$(M,g)$ has no zeroes on that surface, then it has no zeroes anywhere.
 
We now prove Proposition 1. Let $(\Sigma_0,\gamma,\pi)$ be the given
initial surface and initial data for the spacetime Killing vector
fields which generate the $T^2$ group action on $\Sigma_0$. By
definition, $X$ and $Y$ are nonzero everywhere on $\Sigma_0$, and
non parallel everywhere on $\Sigma_0$ as well.  It also follows from
the definition of areal coordinates and the orbital area function $R$
that we have
$$ R^2=\lambda^{-2} \det\left({g(X,X)\quad g(X,Y)}\above0pt {g(Y,X)\quad
    g(Y,Y)}\right)
$$
$$ =\lambda^{-2} \left(g(X,X)g(Y,Y)-g^2(X,Y)\right)\ .  \eqno(5.3)
$$ Note that on $\Sigma_0$,
$$ R^2\mid_{\Sigma_0}=\lambda^{-2}
\left(\gamma(X,X)\gamma(Y,Y)-\gamma^2(X,Y)\right)\ , \eqno(5.4) 
$$ and since $X$ and $Y$ have no zeroes and are nowhere parallel on
$\Sigma_0$, $R^2$ has no zeroes on $\Sigma_0$.
 
Let us suppose that for some point $p\in M$, $R(p)=0$. It follows from
Lemma 1 that there exists a Cauchy surface $\Sigma_p$ such that $X$
and $Y$ are tangent to $\Sigma_p$, and so we have
$$ 0=R^2(p)=\gamma_p(X,X)\gamma_p(Y,Y)-\gamma_p^2(X,Y) \eqno(5.5)
$$ where $\gamma_p$ is the induced spatial metric on $\Sigma_p$. From
(5.5) we see that it must be true that either $X$ or $Y$ is zero at
$p$, or $X$ is parallel to $Y$ at $p$. By Lemma 2, the first
possibility cannot be true. The second possibility is also ruled out
by Lemma 2; we argue this as follows: If $X$ were parallel to $Y$ at
$p$, then we would have $X(p)=\beta Y(p)$. We now consider the Killing
fields $X$ and $Z=X-\beta Y$. On $\Sigma_0$, $X$ and $Z$ have no
zeroes, and are non parallel. But at $p$, $Z(p)=X(p)-\beta Y(p)=0$,
which contradicts Lemma 2.
 
We conclude that $R^2$ and hence $R$ have no zeroes. Since $R$ is
continuous and since we choose the convention $R=+\sqrt{\det g}$, we
have $R>0$ everywhere.
\hfill   $\Box$
 
 Our next result concerns limits of $R$ along past directed paths in
$D^- _{conf}(\gamma,\pi)$, which we recall is the maximal globally
hyperbolic past development of $(\gamma,\pi)$, in which the conformal
components of the metric $(R,U,A,\nu, M_{\!\!\!\!\!_{_1}},
M_{\!\!\!\!\!_{_2}}, G_{\!\!\!\!\!_{_1}}, G_{\!\!\!\!\!_{_2}}) $ exist
as a solution of the Einstein equations (2.8)-(2.10).
 
\bigskip \bf Proposition 2 
 \it For any choice of $T^2$-symmetric initial data $(\gamma,\pi)$,
there exists a unique non-negative number $R_0$ such that every past
inextendible causal path $\Gamma :(-\infty,s_0)\to
D_{conf}^-(\gamma,\pi)$ satisfies \rm
$$ \lim_{s\to -\infty} R\circ\Gamma(s)=R_0 \eqno(5.6)
$$
 
\bigskip \bf Proof: \rm
 There are two cases to consider, depending upon whether\linebreak
\mbox{$D^- _{conf}(\gamma,\pi)\setminus\Sigma_0$} has a past boundary
or not.
 
We first assume that there is no such boundary.  It then follows from
the discussion of Section 4 that $D_{conf}^- (\gamma,\pi)=S^1\times
T^2\times (-\infty,t_0]$, with the conformal coordinates
$(\theta,x,y,t$) covering $D_{conf}^-(\gamma,\pi)$.

Now consider the level sets of $R$ in $D_{conf}^-(\gamma,\pi)$:
$$ \Sigma_\rho:=\{(\theta,x,y,t)\mid
R(\theta,t)=\rho,\quad\rm{with}\quad \rho<\inf_{\Sigma_0}R\}
\eqno(5.7)
$$ We wish to show
 
\bigskip \bf Lemma 2.1  
\it The sets $\Sigma_\rho$, if non-empty, are Cauchy surfaces. \rm
 
\bigskip \bf Proof of Lemma 2.1: \rm 
Since the function $R$ is smooth and since $\nabla R$ is timelike (see
Step 1 of Section 4), it follows that the sets $\Sigma_\rho$ are
smooth, embedded, achronal, spacelike submanifolds of
$D_{conf}^-(\gamma,\pi)$. Since $\rho<\inf_{\Sigma_0}R$, the sets
$\Sigma_\rho$ have no boundary. Thus if we can show that each set
$\Sigma_\rho$ is compact, then it follows from \cite{1} that the
$\Sigma_\rho$'s are Cauchy surfaces.
 
Since they are closed in $D_{conf}^-(\gamma,\pi)$, it suffices to show
that each $\Sigma_\rho$ is bounded. To do this, we note that since a
given $\Sigma_\rho$ is spacelike, we can describe it as the graph of a
function:
$$ \Sigma_\rho=\{(\theta,x,y,f(\theta))\mid\quad\theta\in S^1,(x,y)\in
T^2\} \eqno(5.8)
$$ Now every vector tangent to $\Sigma_\rho$ must be spacelike.
Considering the tangent vector
$$ W=f'\partial_t+\partial_\theta+W_x\partial_x+W_y\partial_y
\eqno(5.9a)
$$ where
$$ W_x=- G_{\!\!\!\!\!_{_1}}- M_{\!\!\!\!\!_{_1}}f' \qquad W_y=-
G_{\!\!\!\!\!_{_2}}- M_{\!\!\!\!\!_{_2}}f' \eqno(5.9b)
$$ we see that $W$ is spacelike -- i.e., $g(W,W)>0$, for $g$ from
(2.7) -- only if $\mid f'\mid<1$. But if $\mid f'\mid<1$, then since
the range of $\theta$ is $2\pi$, we find that the range of $f$ is less
than or equal to $\pi$; that is
$$ \sup_{\Sigma_\rho}t-\inf_{\Sigma_\rho}t\leq \pi \eqno(5.10)
$$ This condition bounds $\Sigma_\rho$, so it must be compact, and
consequently it must be a Cauchy surface.
\hfill 
  $\Delta$

Now consider a past inextendible causal path $\Gamma(s)$,
with $s$, say, in $(-\infty,0]$. Since $R$
monotonically decreases with decreasing $t$, and since $R>0$, it
follows that $\lim_{s\to-\infty}R\circ\Gamma(s)$ exists; we shall call
it $R_0$. If we consider another such path $\hat\Gamma$, we also have
$\lim_{s\to- \infty}R\circ\hat\Gamma(s)$ existing; we call it $\hat
R_0$.
 
Let us assume that $R_0\not=\hat R_0$; without loss of generality,
say \mbox{$R_0<\hat R_0$}. Then for some $p\in D_{conf}^-$, we have
$R(p)=\frac{1}{2}(R_0+\hat R_0)<\hat R_0$. It follows from Lemma 2.1
that $\Sigma_{1/2(R_0+\hat R_0)}$ is a Cauchy surface. But then
$\hat\Gamma$ must intersect $\Sigma_{1/2(R_0+\hat R_0)}$. This is a
contradiction, since we have \mbox{$R\circ\hat\Gamma (s)>\hat R_0$}
for all $s$. So we must
have $R_0=\hat R_0$. This shows that, in this first case, $R$ has the
same limit along all past inextendible causal paths.
 
We now consider the case in which $D_{conf}^-(\gamma,\pi)$ corresponds
to a proper subset of $S^1\times T^2\times(-\infty,t_0]$. Since the
spacetime $D^- _{conf}$ is globally hyperbolic, we know from \cite{17}
that the boundary, which we shall call $\Omega$, is a closed achronal
set. We can also argue that $\Omega$ is bounded and hence compact. To
argue this, we construct a smooth, $T^2$--invariant time function
$\tau$ which goes to zero as one approaches the boundary. (Such a time
function exists, since we know from \cite{12} that one can construct a
smooth time function $\hat\tau$ on $D_{conf}^-$ which approaches
$-\infty$ near $\Omega$; one can then average $\hat\tau$ along $T^2$
and set $\tau=\exp(\hat\tau).)$ So
$\Omega$ is the limit of $\tau=$constant Cauchy surfaces in
$D_{conf}^-$. Now arguing as in Lemma 2.1, we can show that for any
$\tau=$constant Cauchy surface $\Sigma_\tau$, one has a result
equivalent to (5.10):
$$ \sup_{\Sigma_\tau}t-\inf_{\Sigma_\tau}t\leq\pi \eqno(5.11)
$$ where $t$ is the conformal coordinate time. It follows by
continuity in $\tau$ that $\Omega$ must be bounded; compactness
follows.
 
The Whitney Extension Theorem allows one to extend the function $R$
past $D_{conf}^-$; in particular, it guarantees the existence of a
Lipschitz continuous function $\bar R$ on $S^1\times
T^2\times(-\infty,t_0)$ for which $\bar R\mid_{D_{conf}^-}=R$. Now let
us set $\sup_{\Omega}\bar R=R_0.$ The compactness of $\Omega$
guarantees that $R_0$ is attained on $\Omega$.  If $R_0=0$, then it
follows from Proposition 1 and continuity that $\bar R\mid_\Omega=0.$
Hence, since all past inextendible paths in $D_{conf}^-$ must approach
$\Omega$, $R$ must approach the same value -- zero -- along every such
path.
 
It remains to show that $R_0\not=0$ cannot occur. We will show this by
arguing that if it did occur, then one could extend the conformal
coordinate solution of (2.8)-(2.10) into the past of $D_{conf}^-$,
which by presumption cannot be done.
 
So we presume that $R_0\not=0$, and we consider the set of
points
$$ I:=\{(\theta,x,y,t)\in\Omega\mid\bar R(\theta,x,y,t)=R_0\}
\eqno(5.12)
$$ which is nonempty since $\Omega$ is compact. Letting $\xi$ denote
the sup of the time function $t$ on $I$, we locate a point $p\in I$
with $t(p)=\xi$. Such a point exists since $I$ is compact, and we
label its coordinates $(\hat\theta,\hat x,\hat y,\hat
\xi)\leftrightarrow p.$
 
Now it follows from the definition of $p$ that for sufficiently small
$\epsilon>0$, the set
$$ S_{(p,\epsilon)}:=\{(\theta,x,y,\xi)\mid\hat\theta-
\epsilon<\theta<\hat\theta+\epsilon,\quad (x,y)\in T^2\} \eqno(5.13)
$$ is a well defined subset of $D_{conf}^-(\gamma,\pi)\cap\Omega$
which forms a spacelike submanifold and has $R\geq\frac{R_0}{2}$.
Thus, based on our estimates in \mbox{Section 4} with $R$ bounded away
from zero, we can smoothly extend all of the fields $(R,U,A,\nu,
G_{\!\!\!\!\!_{_1}}, G_{\!\!\!\!\!_{_2}}, M_{\!\!\!\!\!_{_1}}$ and $
M_{\!\!\!\!\!_{_2}}\enspace\! )$ to $S_{(p,\epsilon)}$ and set up a
well-posed initial value problem on $S_{(p,\epsilon)}$. Standard
existence results (see, e.g. \cite{3}) then allow us to evolve the
fields to the past of $p$. This contradicts the definition of
$D_{conf}^-$, and thus tells us that $R_0=0$, completing the proof of
Proposition 2.
\hfill  $\Box$

With the uniform limits of the function $R$ along all past
inextendible causal paths in $D_{conf}^-(\gamma,\pi)$ established, one
easily proves the following key part of the proof of Theorem 1.
 
\bigskip \bf Proposition 3 \it Let ($\Sigma_0,\gamma,\pi)$ be
$T^2$-symmetric initial data. Define $R_1$ as $ \inf_{\Sigma_0}R$, and
$R_0$ as the past limit of R along past inextendible paths in
$D_{conf}^-(\gamma,\pi)$. For every $\rho\in(R_0,R_1)$, the $R=\rho$
level set $\Sigma_\rho$ (see equation (5.7)) is a Cauchy surface, and
these $\Sigma_\rho$ foliate the spacetime region $D_{conf}^-
(\gamma,\pi)\cap I^-(\Sigma_{R_1}).$ Further, this spacetime region
admits areal coordinates.\rm
 
\bigskip \bf Proof: \rm
 The proof that $\Sigma_\rho\subset D_{conf}^-(\gamma,\pi)\cap I^-
(\Sigma_{R_1})$ is a Cauchy surface is essentially that given in Lemma
2.1 for a slightly more restricted situation. The generalization here
is that $D_{conf}^- (\gamma,\pi)$ may have a boundary. However, as
shown in Proposition 2, if the boundary $\Omega$ exists, than all past
inextendible causal paths must approach it, with $R$ approaching 0 on
$\Omega$. Hence for $R=\rho>0$ in this case, $\Sigma_\rho$ is bounded
away from the boundary. Compactness then follows; hence as a
consequence of \cite{1}, $\Sigma_\rho$ is a Cauchy surface.
 
Since $R$ is smooth and monotonically decreasing along past causal
path, all values of $R$ between $R_0$ and $R_1$ are realized in order.
Thus we verify that $D_{conf}^-(\gamma,\pi)\cap I^-(\Sigma_{R_1})$ is
foliated by the $\Sigma_R$.
 
Once we have an $R$=const foliation, it readily follows from arguments
of the form surrounding equation (4.15) in \cite{5} that the region
$D_{conf}^- (\gamma,\pi)\cap I^-(\Sigma_{R_1})$ admits areal
coordinates with the metric taking the form (2.3) and satisfying
(2.4)-(2.6).
\hfill  $\Box$
 
We now know that $D_{conf}^-(\gamma,\pi)\cap I^-(\Sigma_{R_1})$ admits
an $R$ foliation and areal coordinates. To finish the proof of Theorem
1 for the contracting direction, it remains to show that
$D_{conf}^-(\gamma,\pi) $ covers the past maximal globally hyperbolic
development $D^-(\gamma,\pi) $ of the initial data. We show this
separately for two different cases: First, in the case that
$t\to-\infty$ (and $R_0\not=0$), and then in the case that $R_0=0$.
 
\bigskip \bf Proposition 4 
\it If $D_{conf}^-(\gamma,\pi)$ has $t\to-\infty$, then
$D_{conf}^-(\gamma,\pi) $ is diffeomorphic to $D^-(\gamma,\pi)$.
 
\bigskip \bf Proof: \rm
Let us for convenience designate the conformal coordinate spacetime
region $D_{conf}^-(\gamma,\pi)$ as $(M,g)$, and the maximal past
development as $(\hat M,\hat g)$. We now suppose that $(M,g)$ is \em
not \rm diffeomorphic to $(\hat M,\hat g)$; then there must be an
isometric embedding
$$ \psi:\quad M\longrightarrow\hat M, \eqno(5.14)
$$ and $\psi(M)$ must have a non empty boundary $\partial\psi(M)$ in
$M$.  Note that it follows from our choice of time convention that
$\partial\psi(M)$ is a past boundary of $\psi(M)$ in $M$.
 
Since the spacetimes $M$ and $\hat M$, as well as the embedding $\psi$
are all $T^2$ invariant, those orbits of the action of $T^2$ which
intersect $\psi(M)$ are contained in $\psi(M)$. We consider a point
$p\in\partial \psi(M)$ and its corresponding $T^2$ orbit $\calO_p
\subset  \partial \psi(M)$. The main idea of the proof is to obtain a
contradiction regarding the causal future $J^+(\calO_p)$ of this orbit
and its intersection with nearby Cauchy surfaces in $\psi(M)$.

We work with two families of Cauchy surfaces. One family is
$\Sigma_t$, where the $\Sigma_t$'s are level sets of the conformal
time coordinate $t$ in $\psi(M)$. The other family is given by
$\Sigma_{\hat t}$, where the $\Sigma_{\hat t}$'s are level sets of any
$T^2$--invariant time coordinate $\hat t$ in $\hat M$ (see Lemma 1.1).
Note that since both time functions are $T^2$--invariant, we have
$t(r)=t(p)$ and $\hat t(r)=\hat t(p)$ for all $r\in \calO_p$, so that
we can refer unambiguously to $t(\calO_p)$ and $\hat t(\calO_p)$.

We use the $\hat t$ time function to prove the following:

\bigskip \bf Lemma 4.1  
\it a) There exists a a sufficiently small $\epsilon > 0$ such that for
$|\hat t - \hat t(\calO_p)|<\epsilon$ we have $J^+(\calO_p)\cap
\Sigma_{\hat t}\ne \Sigma_{\hat t}$ 

b) For any Cauchy surface $\Sigma \subset J^-(\Sigma_{\hat t_1})\cap
\psi(M)$ such that $|\hat t_1- \hat t(\calO_p)|<\epsilon $, one has
$J^+(\calO_p) \cap \Sigma \ne \Sigma$.
 
\bigskip \bf Proof of Lemma 4.1: \rm 
Consider a point $w$ which is contained in the Cauchy surface
$\Sigma_{\hat t(\calO_p)} $, but is not contained in the orbit $
\calO_p$. It follows from $T^2$--invariance that $\calO_p\cap \calO_w$
is empty, and it follows from the acausality of $\Sigma _{\hat
  t(\calO_p)} $ that no path between a point in $\calO_p$ and a point
in $\calO_w$ is causal.

Using the $\hat t$ function we can construct a neighbourhood of $\Sigma _{\hat
  t(\calO_p)} $ which is diffeomorphic to $\Sigma _{\hat
  t(\calO_p)} \times (t(\calO_p)-\delta,t(\calO_p)+\delta)$ for some
$\delta >0$. Let $t_k$ be a monotonically decreasing sequence
converging to $t(\calO_p)$ and let $w_k=(w,t_k)$. If the lemma were
false, then for every value of $k$ there would be a causal path
$\gamma_k$ from $w_k$ to a point in $\calO_p$. By global hyperbolicity
there exists a causal path $\gamma$ at which a subsequence of the
$\gamma_k$'s accumulates. It follows that  $\gamma$ is a causal path
between $\calO_p$ and $\calO_w$, so we have a contradiction, and
point a)  follows.

To establish point b), let us fix a value $\hat t_1$ such that
$J^+(\calO_p)\cap \Sigma_{\hat t_1}\ne \Sigma_{\hat t_1}$, and let
$\Sigma$ be any Cauchy surface for which $\Sigma \subset
J^-(\Sigma_{\hat t_1}) \cap \psi(M)$. Through every point $r\in 
\Sigma_{\hat t_1}$ there is a past--directed causal path which
intersects $\Sigma$. Hence, if $J^+(\calO_p)\cap \Sigma =\Sigma$, then
there would be a past--directed causal path from every point in
$\Sigma _{\hat t_1}$ to a point in $\calO_p$. But we know this is
false, so we must have $J^+(\calO_p)\cap \Sigma \ne \Sigma$. \hfill
$\Delta$

We now use the conformal time $t$ to show that for all Cauchy
surfaces $\Sigma$ contained in $J^-(\Sigma_{\hat t_1}) \cap \psi(M)$
we actually have $J^+(\calO_p)\cap \Sigma = \Sigma$, thus
contradicting Lemma 4.1 if we assume that $M$ is extendible. To do
this, we first note that for any Cauchy surface $\Sigma$ which
intersects $\psi(M)$, it must be true that $\Sigma$ is entirely
contained in $\psi(M)$, and also one must have 
%
%
%
%
$$ \sup_{\hat\Sigma}t-\inf_{\hat\Sigma}t\leq\pi\ , \eqno(5.15)
$$ where $t$ is the conformal coordinate. The proof of these two
claims is essentially a local version of the proof of Lemma 2.1. That
is , starting at a point $p $ in the intersection of $\Sigma$ with
$\psi(M)$, we write $\Sigma$ locally as a graph of a function
$f(\theta)$ (see equation (5.8)), and then note that the spacelike
character of $\Sigma$ implies that $\mid f'\mid<1$. Since the
conformal coordinate time $t$ has a finite value at $p$, and since
$t\to-\infty$ in $\psi(M)$, it follows from $\mid f'\mid<1$ and from
the conformal coordinate metric form (2.7) that $t$ is bounded on
$\Sigma$ and therefore $\Sigma$ cannot escape $\psi(M)$. Then
choosing the graph representation (5.8) of $\Sigma$ globally, one
verifies (5.15) as in Lemma 2.1.
 
We now consider a point $q$ contained in the past set
$J^-(\Sigma,\psi(M))$ such that
$$ \sup_{\Sigma}t-t(q)\geq\pi .  \eqno(5.16a)
$$ Such a point exists, since the compact set $\Sigma$ has finite $t$
and $t\to -\infty$ as one approaches $\partial \psi(M)$. Moreover,
since $t(\calO_q)=t(q) $, we have
$$ \sup_{\Sigma}t-t(\calO_q)\geq\pi .  \eqno(5.16b)
$$
We claim that for any $q$ and corresponding $\calO_q$, the set
$J^+(\calO_q)\cap \Sigma$ is equal to $\Sigma$. To show this, let us
label the coordinates of $q$ as $(\theta_0, x_0, y_0, t_0)$, so that
$(\theta_0,  t_0)$ labels the orbits $\calO_q$ as we vary $q$, and let
us consider the one--parameter family, labeled by $\alpha$, of paths,
labeled by $s$, of $T^2$ orbits 
%
$\{ \Gamma_{\!\!\!\!\!_{_{[\alpha]}}}(s) \mid \alpha\in[- 1,+1],
s>0\}$ given by
$$ \Gamma_{\!\!\!\!\!_{_{[\alpha]}}}(s)=\{(\theta_0+\alpha s, x,
y, t_0+s), \quad x,y\in T^2\}\ . \eqno(5.17a) 
$$
 We
note two important features of this family of paths of $T^2$ orbits:
First, the set of all points in these orbits is contained in the
future set 
$J^+(\calO_q)$. Second,
%
\begin{eqnarray*}
  {\cal S}_{\!\!\!\!\!_{_{[\alpha]}}} & = & \{
  \Gamma_{\!\!\!\!\!_{_{[\alpha]}}}(\pi) \mid\alpha\in[-1,+1]\} \\ & =
  & \{\theta_{\!\!\!\!\!_{_{[\alpha]}}}=\theta_0+\alpha\pi,(x,y)\in
  T^2, t_{\!\!\!\!\!_{_{[\alpha]}}}=\,t_0+\pi \mid\alpha\in[-1,+1]\}
\\
& = & \Sigma_{t_0+\pi} \ . 
\end{eqnarray*}
$$ \eqno(5.18)
$$ 
Hence we have $J^+(\calO_q)\cap\Sigma = \Sigma$.

To finish the proof of Proposition 4, we consider the Cauchy surface
$\Sigma_{\hat t_1}$, defined in Lemma 4.1. Since $\Sigma_{\hat t_1}$
intersects $\psi(M)$, it must be contained in $\psi(M)$ and hence
$T:=\inf_{\Sigma_{\hat t_{1}}}t$ is finite. It then follows from Lemma
4.1 that for $t<T$ we have $J^{+}(\calO_p)\cap \Sigma_t\ne \Sigma_t$.

Now let us consider a sequence of points $p_i$ such that $p_i \in
J^{-}( \Sigma_{\hat t_{1}})\cap \psi(M)$, $p_{i+1}\in I^-(p_i)$, $p_i\in
  J^+(p)$, and $p_i\to p$, where $p$ is the point we have chosen on
  $\partial \psi (M)$. Since the points $p_i$ do not have an
  accumulation point in $\psi(M)$ we have
  $\lim_{i\to\infty}t(p_i)=-\infty$. This tells us that for any fixed
  $i$ there is some $i_\pi$ such that if $j> i_\pi$ then $t(p_j)<
  t(p_i)-\pi$. It then follows, as shown above, that
  $J^+(\calO_{p_j})\cap \Sigma_{t_{p_i}}= \Sigma_{t_{p_i}}$. Since
  $t(p_i)<T$, we have a contradiction, from which it follows that $\psi
  (M) = \hat M$. \hfill $\Box$

%
%

Proposition 4 shows that $D^-_{conf}(\gamma,\pi)\approx
D^-(\gamma,\pi)$ if $t\to-\infty$ in $D^-_{conf}$. To show that
$D^-_{conf}(\gamma,\pi)\approx D^- (\gamma,\pi)$ if  $R_0=0$, we
rely on a more general result, part of which we will need to complete
the proof for the expanding direction.
 
\bigskip \bf Proposition 5 \rm
 Let $(M,g)$ be any globally hyperbolic development of $T^2$-symmetric
initial data. If $R\to 0$ along every past inextendible causal path,
then ($M,g)$ covers the past maximal development of ($\gamma,\pi$). If
$R\to\infty$ along every future inextendible causal path, then $(M,g$)
covers the future maximal development of $(\gamma,\pi)$.
 
\bigskip \bf Proof: \rm
 If ($M,g)$ is not maximal to the past, then there exists a globally
hyperbolic spacetime $(\hat M,\hat g)$, with a smooth proper embedding
$\psi(M)\subset\hat M$, and with $\partial\psi(M)$ containing a
portion $\partial^- \psi(M)$ to the past of $\psi(M)$. Let
$p\in\partial^-\psi(M)$. If $R\to0$ along every past inextendible
causal path, then $R(p)=0.$ This disagrees with Proposition 1, so
$(M,g)$ must be maximal to the past.
 
If $(M,g$) is not maximal to the future, then similarly there is a
spacetime ($\check M,\check g)$ with a smooth proper embedding
$\psi(M)\subset \check M$, and with a portion $\partial^+\psi(M)$ of the
boundary to the future of $\psi(M).$ Letting $q\in\partial^+\psi(M)$,
we see that if $R\to\infty$ along every future inextendible causal
path, then $R$ blows up near $q$. But since the Killing vector fields
in ($\check M,\check g)$ are at least $C^1$ everywhere, it follows
that $R$ must be bounded everywhere. Hence $(M,g)$ must be maximal to
the future.
\hfill 
  $\Box$
 
\section{Analysis in the Expanding direction} 
 
The Proof of Theorem 1 for the expanding -- $R$ increasing --
direction is more direct than for the contracting direction, since we
work directly with the areal coordinate components of the metric (see
equation(2.3)) To do this, we need to start with data on an
$R$=constant Cauchy surface. Let
$R_0$ and $R_1$ be as in Proposition 3; then
as shown in Sections 4-5, such surfaces
exist for all $R\in(R_0,R_1)$ --- they lie to the past of the initial
surface $\Sigma_0$ with data $(\gamma,\pi)$ from the hypotheses of
Theorem 1. Let us pick one such surface, with say $R=R_2$. The
spacetime $D^-(\gamma,\pi)$ induces initial data for the areal
component fields ($U,A,\nu,\alpha, G_{\!\!\!\!\!_{_1}},
G_{\!\!\!\!\!_{_2}}, M_{\!\!\!\!\!_{_2}\enspace\! },
M_{\!\!\!\!\!_{_1}})$ on $\Sigma_{t_2=R_2}$, and we have local
existence for the initial value problem for these fields. To prove
global existence -- i.e. to show that we can evolve the fields
$(U,A,\nu,\alpha, G_{\!\!\!\!\!_{_1}}, G_{\!\!\!\!\!_{_2}},
M_{\!\!\!\!\!_{_2}\enspace\! }, M_{\!\!\!\!\!_{_1}})$ via the Einstein
equations (2.4)-(2.6) to $t\to\infty$ -- what we need to do (as shown
in \cite{14}) is prove that for any finite interval $[t_2,T)$ on which
they exist as a solution to (2.4)-(2.6), these functions are uniformly
$C^2$ bounded. Again, we do this in a series of steps.
 
\bigskip \underline{Step 1} (Bounds on $\alpha,\nu$ and $\alpha_t$)
 
As in Section 4, we use light cone estimates here to establish the
bounds we need for the fields $U$ and $A$ and their derivatives.
However, since the wave equations (2.5a) and (2.5b) for $U$ and $A$
involve $\alpha$ and its derivatives, and since the constraint
equation (2.4c) for $\alpha$ involves $\nu$, we first need to bound
these quantities. The first step towards doing this is an energy
monotonicity result:
 
Let us define
$$ {\cal E}(t):=\int\limits_{S^1}[\alpha^{-\frac{1}{2}}U_t^2+
\alpha^{\frac{1}{2}}U_\theta^2+\frac{e^{4U}}{4t^2}
(\alpha^{-\frac{1}{2}}A_t^2+\alpha^{\frac{1}{2}}A_\theta^2)]d\theta.
\eqno(6.1)
$$ Using equations (2.4)-(2.6), and integration by parts, we calculate
$$
\frac{d}{dt}{\cal E}=\frac{-K^2}{t^3}\int\limits_{S^1}\frac{1}{2\alpha}
[\alpha^{-\frac{1}{2}}U_t^2+
\alpha^{\frac{1}{2}}U_\theta^2+\frac{e^{4U}}{4t^2}
(\alpha^{-\frac{1}{2}}A_t^2+\alpha^{\frac{1}{2}}A_\theta^2)]
e^{2\nu}\alpha^2d\theta
$$
$$
-\frac{1}{2t^3}\int\limits_{S^1}(e^{4U}\alpha^{\frac{1}{2}}A_\theta^2)
d\theta
-\frac{2}{t}\int\limits_{S^1}(\alpha^{-\frac{1}{2}}U_t^2)d\theta <0
\eqno(6.2)
$$ This shows that ${\cal E}(t)$ decreases monotonically in $t$. So
in particular, we have
$$ {\cal E}(t)<{\cal E}(t_2) \eqno(6.3)
$$ for all $t>t_2$.
 
Now we consider the quantity
$$ \tilde\nu:=-\nu-\frac{1}{2}\ln\alpha \eqno(6.4)
$$ The spatial derivative of $\tilde\nu$, as a consequence of the
constraint (2.4b), is given by
$$ \tilde\nu_\theta=-2tU_tU_\theta-\frac{e^{4U}}{2t}A_tA_\theta
\eqno(6.5)
$$ It follows readily from the definition of ${\cal E}(t) $ -- and
from the algebraic fact that, for any $a,b$ and $c>0$, one has $\mid
ab\mid\leq\frac{1}{2c}a^2+2cb^2$ -- that
$$ \int\limits_{S^1}\mid\tilde\nu_\theta\mid
d\theta\leq t {\cal E}(t).  \eqno(6.6)
$$ Hence, using the monotonicity of ${\cal E}(t)$, we find that for
all $t\geq t_0$,
$$ \int\limits_{S^1}\mid\tilde\nu_\theta\mid d\theta\leq
t{\cal E}(t_2).  \eqno(6.7)
$$
 
As a consequence of (6.7) and the mean value theorem, we can control
the variance of $\tilde\nu$ on a given Cauchy surface at areal time
$t$. That is, for any $\theta_1,\theta_2\in S^1$, and for any
$t\in[t_2,T)$, we have
\begin{eqnarray*}
  \mid\tilde\nu(\theta_2,t)-\tilde\nu(\theta_1,t)\mid \quad & = &
  \quad\mid\int\limits_{\theta_1}^{\theta_2}\tilde\nu_\theta
  d\theta\mid \\ & \leq &
  \int\limits_{\theta_1}^{\theta}\mid\tilde\nu_\theta\mid d\theta \\ &
  \leq & \int\limits_{S^1}\mid\tilde\nu_\theta\mid d\theta\leq
  t{\cal E}(t_2).
\end{eqnarray*}
$$ \eqno(6.8)
$$

We calculate the time derivative of $\tilde\nu$ from constraints
(2.4a) and (2.4c); we get
$$ \tilde\nu_t=-t[U_t^2+\alpha U_\theta^2+\frac{e^{4U}}{4t^2}(A_t^2+
A^2_\theta)]+\lambda \frac{e^{2\nu}}{4t^3}\alpha K^2.  \eqno(6.9)
$$ It follows immediately from (6.9) that we have two inequalities for
$\tilde\nu_t$:
$$ \tilde\nu_t\geq -t[U_t^2+\alpha
U_\theta^2+\frac{e^{4U}}{4t^2}(A^2_t+ \alpha A_\theta^2)] \eqno(6.10a)
$$ and
$$\tilde\nu_t\leq \lambda \frac{e^{-2\tilde\nu}}{4t^3}K^2.  \eqno(6.10b)
$$ Using (6.10a), we obtain
\begin{eqnarray*} 
  \int\limits_{S^1}\tilde\nu d\theta & = & \int\limits_{t_2}^t
  \frac{d}{dt} \left(\int\limits_{S^1}\tilde\nu d\theta\right)d\tau \\ 
  & = &
  \int\limits^t_{t_2}\left(\int\limits_{S^1}\tilde\nu_td\theta\right)d\tau
  \\ & \geq & -\int\limits_{t_2}^t\tau\int\limits_{S^1}[U_t^2+\alpha
  U_\theta^2+ \frac{e^{4U}}{4\tau^2}(A_t^2+\alpha A^2_\theta)]d\theta
  d\tau \\ & = & -\int\limits_{t_2}^t\tau{\cal E}(\tau)d\tau \\ &
  \geq & -\int\limits_{t_2}^t\tau{\cal E}(t_2)d\tau \\ & \geq &
  -{\cal E}(t_2)(t^2-t_2^2)/2\ ,
\end{eqnarray*}
$$ \eqno(6.11)
$$ which controls $\int_{S^1}\tilde\nu d\theta$ from below. Then if we
combine (6.11) and (6.8), we derive a lower bound on $\tilde\nu$
itself. We do this as follows: If we apply (6.11) to the left hand
side of the identity
$$
\int\limits_{S^1}\tilde\nu=\int\limits_{S^1}[\min_{S^1}\tilde\nu+\tilde\nu
-\min_{S^1}\tilde\nu] \eqno(6.12)
$$ we obtain
$$ -{\cal E}(t_2)(t^2-t_2^2)\leq2\pi\min_{S^1}\tilde\nu+
\int\limits_{S^1}(\tilde\nu-\min_{S^1}\tilde\nu).  \eqno(6.13)
$$ Applying (6.8) to the second term on the right hand side of (6.13),
we get
$$ -{\cal E}(t_2)(t^2-t_2^2)\leq2\pi\min_{S^1}\tilde\nu+ct
\eqno(6.14)
$$ for some constant $c$. Rearranging (6.14), we have
$$ \min_{S^1}\tilde\nu\geq c(t) \eqno(6.15)
$$ where $c(t)$ is a bounded function of $t$ on $[t_2,T).$
 
To obtain an upper bound for $\tilde\nu$, we use (6.10b), together
with (6.15).  Specifically, we have
\begin{eqnarray*} 
  \tilde\nu_t\ & \leq & \lambda \frac{e^{-2\tilde\nu}}{4t^3}K^2 \\ & \leq &
 \lambda  \frac{e^{-2\min_{S^1}\tilde\nu}}{4t^3}K^2 \\ & \leq &
  \lambda \frac{e^{-2c(t)}}{4t^3}K^2
\end{eqnarray*}
$$ \eqno(6.16)
$$ from which it follows that $\tilde\nu_t$ is controlled into the
future. Control of $\tilde\nu$ immediately follows; so we have upper
and lower bounds for $\tilde\nu$ on $S^1\times[t_2,T)$.
 
We now use the bounds just established for $\tilde\nu$ to obtain
controls for $\nu$ and $\alpha$. We start by noting that the
constraint (2.4b) together with the definition of $\tilde\nu$ leads to
the expression
$$ \partial_t(\ln\alpha)=-\frac{1}{2}\lambda \frac{e^{-2\tilde\nu}}{t^3}K^2
\eqno(6.17)
$$ With $\tilde\nu$ bounded above and below on $S^1\times[t_2,T)$, it
follows from (6.17) that $\partial_t(\ln\alpha)$ -- and hence
$\ln\alpha$ and $\alpha$ -- are as well. Since
$$ \nu=-\tilde\nu - \frac{1}{2} \ln\alpha \eqno(6.18)
$$ we thus obtain bounds for $\nu$. Finally, as a consequence of these
bounds on $\nu$ and $\alpha$, equation (2.4c) tells us that $\alpha_t$
is bounded.
 
This analysis does not lead to bounds for $\alpha_\theta$ on
$S^1\times[t_2,T)$.  While this might appear, from equations (2.5a)
and (2.5b) to be a potential obstacle to using light cone estimates
for $U$ and $A$, we will see in the next step that it is not.
 
\bigskip \underline{Step 2} (Bounds on $U,U_\theta,U_t$ and $
A,A_\theta,A_t$)
 
As with the contracting direction, to obtain the light cone estimates
we use for bounding $U$ and $A$ and their derivatives in the expanding
direction, we find it useful to treat $U$ and $A$ as components of a
wave map $\phi$. The base geometry and target geometries are
different, but the idea is very much the same.  We take for the base
geometry ($S^1\times{\cal R},{}^2g)$ with the (non- flat) Lorentz
metric
$$ {}^2g=-dt^2+\frac{1}{\alpha}d\theta^2 \ \ , \eqno(6.19)
$$ and for the family of target geometries we use
(${\cal R}^2,h_{(t)})$, with \mbox{$(t$-dependent)} Riemannian
metrics
$$ h_{(t)}=dU^2+\frac{e^{4U}}{4t^2}dA^2 \eqno(6.20)
$$ The maps take the component form
\begin{eqnarray*} 
  \phi:\quad S^1\times{\cal R}^1 & \longrightarrow & {\cal R}^2 \\ 
  (\theta,t) & \longmapsto & \phi^a(\theta,t)=\left({U(\theta,t)}
    \above0pt {A(\theta,t)} \right).
\end{eqnarray*}
$$ \eqno(6.21)
$$
 
Since the base geometry is not flat, the metric-compatible\footnote{As
  in Section 4, $D$ is not fully metric-compatible as a consequence of
  the $t$-dependence of $h_{(t)}$ in (6.20)} covariant derivative $D$
for these maps has non vanishing Christoffel coefficients
$$ \Gamma^t_{\enspace tt}=0\qquad\Gamma^t_{\enspace t\theta}=0 \qquad
\Gamma^t_{\enspace\theta\theta}=-\frac{\alpha_t}{2\alpha}
$$
$$ \Gamma^\theta_{\enspace\theta\theta}=-\frac{\alpha_\theta}{2\alpha}
\qquad\Gamma^\theta_{\enspace\theta t}=-\frac{\alpha_t}{2\alpha}\qquad
\Gamma^\theta_{tt}=0, \eqno(6.22)
$$ as well as target Christoffel coefficients
$$ \Gamma^U_{\enspace UU}=0\qquad\Gamma^U_{\enspace UA}=0 \qquad
\Gamma^U_{\enspace AA}=-\frac{e^{4U}}{2t^2}
$$
$$ \Gamma^A_{\enspace AA}=0 \qquad\Gamma^A_{\enspace A U}=2\qquad
\Gamma^A_{UU}=0. \eqno(6.23)
$$ The wave equation for $\phi=\left( U \above0pt A \right)$ now takes
the form
$$ \Box\phi^a=\psi^a \eqno(6.24a)
$$ where
$$ \psi^a=\left({\frac{U_t}{t}} \above0pt {-\frac{A_t}{t}}\right).
\eqno(6.24b)
$$ We have the corresponding energy-momentum tensor, defined as in
(4.15), taking the form
$$ T_{\mu\nu}=U_\mu U_\nu+\frac {e^{4U}}{4t^2}A_\mu A_\nu+
\frac{1}{2}g_{\mu\nu}[U_t^2-\alpha U_\theta^2+
\frac{e^{4U}}{4t^2}(A_t^2-A_\theta^2)].
$$ Since the base metric ${}^2g$ is not flat, while we can readily
choose a pair of everywhere independent null vector fields
$l=\frac{1}{\sqrt{2}}(\partial_t+ \alpha^{1/2}\partial_\theta)$ and
$n=\frac{1}{\sqrt{2}}(\partial_t- \alpha^{1/2}\partial_\theta)$ for
the base, these do not generally define global null coordinates. The
light cone argument does not really need such coordinates, however. It
is sufficient to work with $l,n$ and their integral paths, which are
well-behaved since $\alpha$ is bounded.
 
Calculating as in equations (4.17)-(4.18), we derive from (6.25) and
(6.22)-(6.23) the equations
$$ n(T_{ll})=-\frac{\alpha_t}{\sqrt{2}\alpha}T_{ll}-\frac{1}{\sqrt{2}}
(\frac{U_t^2}{t}+\frac{e^{4U}}{4t^3}\alpha A_\theta^2+
\frac{\alpha^{1/2}}{t}U_\theta U_t+\alpha^{1/2}\frac{e^{4U}}{4t^3}
A_\theta A_t) \eqno(6.25a)
$$ and
$$ l(T_{nn})=-\frac{\alpha_t}{\sqrt{2}\alpha}T_{nn}-\frac{1}{\sqrt{2}}
(\frac{U_t^2}{t}+\frac{e^{4U}}{4t^3}\alpha A_\theta^2-
\frac{\alpha^{1/2}}{t}U_\theta U_t-\alpha^{1/2}\frac{e^{4U}}{4t^3}
A_\theta A_t).  \eqno(6.25b)
$$
 
It is important here that while $\alpha$ and $\alpha_t$ appear in
equations (6.25), $\alpha_\theta$ does not. Thus we find that the
right hand side of (6.27) involves only terms quadratic in
$U_\theta,U_t,\frac{e^{2U}}{2t}A_\theta$ and $\frac{e^{2U}}{2t}A_t$,
with the coefficients of these terms all bounded on $S^1\times[t_2,T)$
as a consequence of the estimates from Step 1. Noting this, and
writing $T_{ll}=T_{tt}+\alpha^{1/2}T_{\theta t}$ and $T_{nn}=T_{tt}-
\alpha^{1/2}T_{\theta t}$, we see that if we set $E=T_{tt}$,
$P=\alpha^{1/2}T_{t\theta}$, and $J=$(right hand side of (6.25a)), and
$L=$(right hand side of (6.25b)), then (6.25) matches (4.6).
 
Since (6.25) has been determined to have the appropriate form, we see
that the rest of the light cone estimate argument goes through more or
less as discussed in Step 2 of Section 4, from equation (4.19) on. We
need to replace null coordinates by null paths generated by the null
vector fields $n$ and $l$, but this does not affect the argument
significantly. We thus find that for any point $(\theta,t)\in
S^1\times(t_2,T]$, we have $E(\theta,t)$ bound by data on the initial
hypersurface $\Sigma_{t_2}$. It immediately follows that $U$ and $A$
are uniformly $C^1$ bounded on $S^1\times(t_2,T)$.
 
\bigskip \underline{Step 3} (Bounds on
$\nu_t,\alpha_\theta,\nu_\theta,\alpha_{\theta t}$ and $\alpha_{tt})$
 
Since the constraint equation (2.4a) expresses $\nu_t$ in terms of
$U_t,U_\theta$, $U,A_t,A_\theta,\nu$ and $t$; and since we have shown
(Steps 1 and 2) that these are all bounded, we immediately obtain from
(2.4a) bounds on $\nu_t$ as well.
 
The same argument does not work for $\nu_\theta$, since equation
(2.4b) involves $\alpha_\theta$, for which we do not yet have bounds.
However, we recall from equation (5.5) that the expression for
$\tilde\nu_\theta$ involves only $t,U_t,U_\theta,A_t,A_\theta$ and
$U$, so $\tilde\nu_\theta$ must be bounded. Then, if we write (2.4c)
in the form
$$ \alpha_t=-\lambda \frac{e^{-2\tilde\nu}}{t^3}\alpha K^2 \eqno(6.26)
$$ and calculate the $\theta$ derivative of both sides (and use local
smoothness), we obtain
$$
\partial_t\alpha_\theta=(-\lambda \frac{e^{-2\tilde\nu}}{t^3}K^2)\alpha_\theta+
(2\lambda \frac{e^{-2\tilde\nu}}{t^3}K^2\tilde\nu_\theta\alpha).  \eqno(6.27)
$$ Since the quantities in parentheses are controlled, we may
integrate this differential equation for $\alpha_\theta$ 
in time, and thereby obtain bounds for $\alpha_\theta.$
 
It then follows from the relation $\nu_\theta=-
\tilde\nu_\theta-\frac{1}{2}\frac{\alpha_\theta}{\alpha}$ that
$\nu_\theta$ is bounded.
 
We now have uniform $C^1$ bounds on all of the primary fields
$U,A,\nu$ and $\alpha$. Equation (6.27) tells us that $\alpha_{\theta
  t}$ is also bounded, and if we calculate the time derivative of
equation (2.4c), we get
$$ \alpha_{tt}=\lambda \frac{e^{2\nu}}{t^3}\alpha K^2 (-2\alpha\nu_t-
2\alpha_t+\frac{3}{t}) \eqno(6.28)
$$ which implies that $\alpha_{tt}$ is bounded as well.
 
To go any further, we need to use light cone estimates again.
 
\bigskip \underline{Step 4} (Bounds on Second Derivatives)
 
If we take time derivatives of the wave equations (2.5) for $U$ and
$A$, then we get wave equations for $U_t$ and $A_t$, which we can
write in the wave map form
$$ \Box
U_t=-\frac{(U_t)_t}{t}+[\frac{1}{2}\alpha_{tt}-\frac{\alpha_t^2}{4\alpha}
-\frac{e^{4U}}{t^2}\alpha A_\theta^2+\frac{1}{t^2}]U_t
$$
$$ +(\frac{e^{4U}}{t^2}\alpha A_\theta
A_t)U_\theta-\alpha\frac{e^{4U}}{2t^3} A_\theta^2 \eqno(6.29a)
$$
$$ \Box
A_t=\frac{(A_t)_t}{t}+[\frac{1}{2}\alpha_{tt}-\frac{\alpha_t^2}{4\alpha}
-4\alpha U_\theta^2-\frac{1}{t^2}]A_t
$$
$$ +4\alpha U_\theta U_t A_\theta ; \eqno(6.29b)
$$ jointly, we have
$$ \Box\phi^a_t= \psi_{\!\!\!\!\!_{_1}}^a. \eqno(6.30)
$$ The two components of $\psi_{\!\!\!\!\!_{_1}}^a$ correspondent to
the right hand sides of (6.29).
 
The important thing to note is that all of the quantities in
$\psi_{\!\!\!\!\!_{_1}}^a$ except $U_{tt}$ and $A_{tt}$ have been
shown in previous steps to be controlled. More importantly, we find
that we may infer from (6.29) that the quantities
$$ E_{\!\!\!_{_1}}=\frac{1}{2}U_{tt}^2+\frac{1}{2}\alpha
U^2_{t\theta}+
\frac{e^{4U}}{4t}(\frac{1}{2}A_{tt}^2+\frac{1}{2}\alpha A^2_{t\theta})
\eqno(6.31a)
$$
$$
P_{\!\!\!_{_1}}=U_{tt}U_{t\theta}+\frac{e^{4U}}{4t^2}A_{tt}A_{t\theta}
\eqno(6.31b)
$$ satisfy equations of the form
$$ n(E_{\!\!\!_{_1}}+ P_{\!\!\!_{_1}})= J_{\!\!\!_{_1}} \eqno(6.32a)
$$
$$ l(E_{\!\!\!_{_1}}- P_{\!\!\!_{_1}})= L_{\!\!\!_{_1}} \eqno(6.32b)
$$ where $ J_{\!\!\!_{_1}}$ and $ L_{\!\!\!_{_1}}$ involve nothing but
controlled quantities together with terms quadratic in $U_{tt},U_{t
  \theta},A_{tt}$ and $A_{t\theta}$.
 
Hence we may repeat the light cone estimate argument as in \mbox{Step
  2} and thereby verify that $U_{tt},U_{t\theta},A_{tt}$ and
$A_{t\theta}$ are all bounded on $S^1\times[t_0,T)$. Further, using
the wave equations (2.5a)-(2.5b), we get bounds on $U_{\theta\theta}$
and $A_{\theta\theta}$; then, using arguments of the sort discussed in
\mbox{Step 4}, we obtain $C^2$ bounds on $\nu$ and $\alpha$ as well.
Thus we have uniform $C^2$ bounds on all of the primary fields.
 
One could repeat this ``boot strap" type argument step-by-step and
obtain bounds on higher order derivatives. However, $C^2$ bounds are
sufficient for the theorems we cite \cite{14} to establish global
existence, so we have proven existence for $t\to\infty$ of the
variables $U,A,\alpha$ and $\nu$.
 
\bigskip \underline{Step 5} (Extension of the Shift Functions)

It remains to show that the shift functions $ G_{\!\!\!\!\!_{_1}},
G_{\!\!\!\!\!_{_2}}, M_{\!\!\!\!\!_{_1}} $ and $ M_{\!\!\!\!\!_{_2}}$
extend to $R=t\to\infty$. Since the constraint equations (2.6) for
these functions in areal coordinates are essentially identical to the
constraint equations (2.10) for them in conformal coordinates, the
procedure for proving  that they extend is 
just that found in Step 6 of Section 2.
 
\bigskip This completes our proof that the areal coordinate
development of the initial data on $\Sigma_{t_2}$ extends to
$R\to\infty$. As shown in Proposition 5, if $R\to\infty$ in the future
development of a set of initial data, then that development must be
maximal. Hence we have completed the proof of our main result, Theorem
1.
 
\section{Conclusions} 
 
The primary motivation for this work has been to set up a framework --
including a geometrically based time foliation -- for studying Strong
Cosmic Censorship and other global issues in a family of spacetimes
which is larger and more complicated than the Gowdy spacetimes, but
still can be studied via (1+1)- dimensional PDE analysis. Indeed, by
relaxing the Gowdy requirement that the twist quantities be non zero,
one obtains field equations which are considerably more intricate than
in the Gowdy case\footnote{The Gowdy equations in areal coordinates
  are the same as (2.4)-(2.5), with $K=0$ and $\alpha=1$.}. Hence the
tools we are developing in working with the $T^2$-symmetric spacetimes
could be more generalizable than those developed in working with Gowdy
spacetimes.
 
Our work here obtains this framework. While the $R=t$ foliation we have
obtained here should -- as evidenced by its importance in the Gowdy
spacetimes -- prove to be very useful, there is another
geometrically-based foliation of considerable interest for these and
other spacetimes: the constant mean curvature (CMC) foliation. Rendall
\cite{18} has studied CMC foliations on $T^2$-symmetric spacetimes --
for the Einstein-Vlasov and Einstein-wave map equations as well as for
the Einstein vacuum equations-- and he has shown that if such a
spacetime admits at least one CMC Cauchy surface $\Sigma$, then it
admits a CMC foliation from a neighborhood of $\Sigma$ back to the
singularity. We expect that 
our areal coordinate foliation provides the barriers necessary to guarantee
the existence of the needed first CMC Cauchy surface.  We also expect
that one might 
be able to use our result to show that the CMC foliation covers the
entire maximal domain of dependence of the spacetime.
 
{\bf Acknowledgments} 
J.I. wishes to thank Alan Rendall for useful discussions. Various
portions of this work were carried out at the Institute for
Theoretical Physics in Santa Barbara, at the Erwin Schr\"odinger
Institute in Vienna, and at the Max Planck Institute for Gravitational
Physics in Potsdam; we thank all these institutions. One of us (J.I.)
also thanks the University of Tours for hospitality while a portion of
this paper was being written. Partial support for this research has
come from  NSF grants { PHY-9507313} at Oakland, PHY-9308117  at
Oregon, and PHY-9503133  at Yale.
 The research of P.T.C. was supported in part by a Polish
  Research Council grant KBN 2P302 095 06.
 
 
\end{document}